\documentclass[a4paper,11pt]{article}
\pdfoutput=1 

\usepackage{jinstpub} 
\usepackage[dvipsnames]{xcolor}

\usepackage[american]{babel}

\usepackage{amsmath}
\usepackage{subcaption}
\usepackage{booktabs}
\usepackage{graphicx}
\usepackage{dcolumn}
\usepackage{bm}

\newcommand{\q}[2]{\ensuremath{#1\ \mathrm{#2}}}

\newcommand{\AAxmax}{\ensuremath{A_{x, m}^2}}
\newcommand{\AAymax}{\ensuremath{A_{y, m}^2}}
\newcommand{\Axmax}{\ensuremath{A_{x, m}}}
\newcommand{\Aymax}{\ensuremath{A_{y, m}}}
\newcommand{\neff}{\ensuremath{n_{\mathrm{eff}}}}
\newcommand{\thscreen}{\ensuremath{\theta_{\mathrm{screen}}}}
\newcommand{\thxymax}{\ensuremath{\theta_{x,y,\mathrm{max}}}}
\newcommand{\thyeff}{\ensuremath{\theta_{y,\mathrm{eff}}}}

\title{Experimental 3-dimensional tracking of the dynamics of a single
  electron in the Fermilab Integrable Optics Test Accelerator (IOTA)}

\author[a]{Aleksandr~Romanov,}
\author[a]{James~Santucci,}
\author[a]{Giulio~Stancari,}
\author[a]{Alexander~Valishev,}
\author[b]{and Nikita~Kuklev}
\affiliation[a]{Fermi National Accelerator Laboratory, Batavia, IL 60510,
  USA}
\affiliation[b]{The University of Chicago, Department of Physics,
  Chicago, IL 60637, USA}
\emailAdd{aromanov@fnal.gov}

\arxivnumber{2012.04148} 

\proceeding{Special Issue on IOTA Beam Physics\\
  27 September 2021}

\abstract{
  We present the results of experimental studies on the transverse and
  longitudinal dynamics of a single electron in the IOTA storage
  ring. IOTA is a flexible machine dedicated to beam physics
  experiments with electrons and protons. A method was developed to
  reliably inject and circulate a controlled number of electrons in
  the ring. A key beam diagnostic system is the set of sensitive
  high-resolution digital cameras for the detection of synchrotron
  light emitted by the electrons. With 60--130 electrons in the
  machine, we measured beam lifetime and derived an absolute
  calibration of the optical system. At exposure times of 0.5~s, the
  cameras were sensitive to individual electrons. Camera images were
  used to reconstruct the time evolution of oscillation amplitudes of
  a single electron in all 3~degrees of freedom. The evolution of
  amplitudes directly showed the interplay between
  synchrotron-radiation damping, quantum excitations, and scattering
  with the residual gas. From the distribution of measured
  single-electron oscillation amplitudes, we deduced transverse
  emittances, momentum spread, damping times, and beam
  energy. Estimates of residual-gas density and composition were
  calculated from the measured distributions of vertical scattering
  angles. Combining scattering and lifetime data, we also provide an
  estimate of the aperture of the ring. To our knowledge, this is the
  first time that the dynamics of a single electron are tracked in all
  three dimensions with digital cameras in a storage ring.
}

\begin{document}
\maketitle
\flushbottom

\section{Introduction}

Observation of a single electron in storage rings has a long history
that goes back to experiments at AdA, the first electron-positron
collider~\cite{bernardini2004ada, BonolisPancheriAdA}. Observation of
discrete steps in radiation intensity offers unique metrology
capabilities associated with the absolute calibration of circulating
currents and radiation properties~\cite{RIEHLE1988262, BRANDT2007445,
  Klein_2010:BESSY}. Another set of experiments was focused on
measurements of synchrotron oscillations by registering deviations of
photon arrival times with respect to the revolution reference
signal~\cite{PINAYEV199417, ALESHAEV199580, PINAYEV199671}. A decade
ago, advancements in digital imaging technology allowed experimenters
to obtain digital images of radiation from single circulating
electrons, but exposure times were too long to resolve and track
instantaneous oscillation amplitudes~\cite{Koschitzki:2010zz}.

This paper presents the results of a first series of experiments
dedicated to a systematic study of an electron's dynamics in the
longitudinal and transverse planes by analyzing high-resolution
digital images obtained with sensitive cameras. The experiments were
carried out in March~2019 during IOTA Run~1. Studies on the precise
measurement of photon arrival times will be presented separately.

Section~\ref{sec:iota} of this paper briefly describes the IOTA
storage ring. Sections~\ref{sec:reconstruction}
and~\ref{sec:statistics} cover the amplitude reconstruction method and
the statistical properties of the observed
parameters. Section~\ref{sec:results} presents the experimental
results: the evolution in time of the amplitude of an electron's
trajectory; single particle emittances and damping times; the lifetime
of a beam with about 100~electrons; description of residual gas
properties; and an estimate of ring acceptance. The last section
(Section~\ref{sec:conclusions}) concludes the paper with a summary and
discussion of the main results.

\section{The IOTA Storage Ring}
\label{sec:iota}

\begin{figure}
  \centering
  \includegraphics[width=0.90\textwidth]{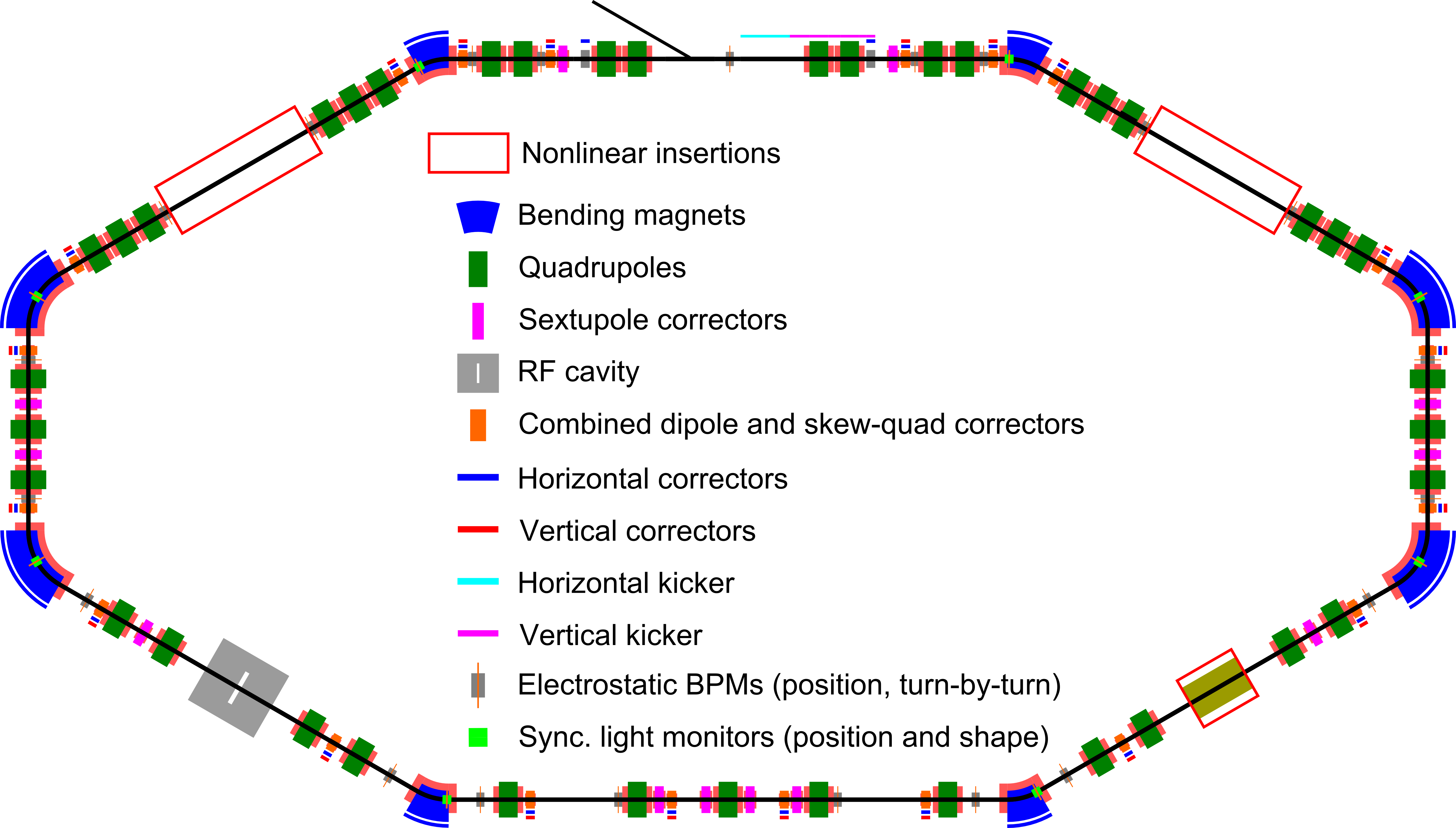} \\
  \includegraphics[width=0.90\textwidth]{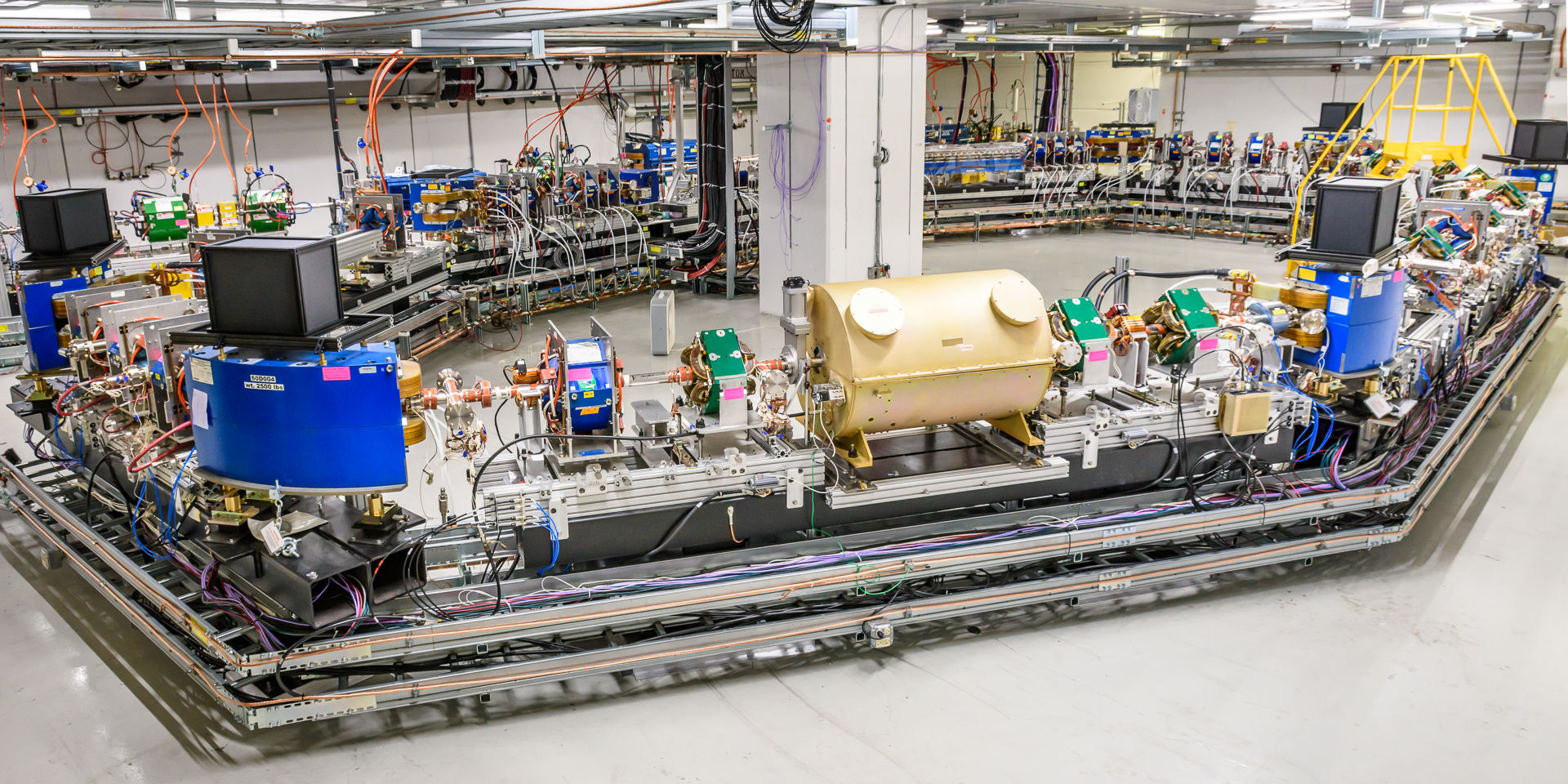}
  \caption{IOTA schematic layout (top) and photograph (facing towards
    the RF cavity).}
  \label{fig:IOTA_view}
\end{figure}

The Integrable Optics Test Accelerator (IOTA) was recently
commissioned with 100~MeV/$c$ electrons as part of the Fermilab
Accelerator Science and Technology (FAST) facility~\cite{IOTA:JINST,
  romanov2020recent}. IOTA is a storage ring with a circumference of
40~m (Figure~\ref{fig:IOTA_view}). It can store electron or proton
beams at momenta between 50 and 150~MeV/$c$ and it can be reconfigured
to accommodate different experiments. The main goal of IOTA is to
demonstrate the advantages of nonlinear integrable lattices for
high-intensity beams and to demonstrate new beam cooling
methods~\cite{DN:PRSTAB, IOTA:JINST}.

Low-emittance and highly configurable electron bunches are injected
from the FAST superconducting linac~\cite{garbincius2013proposal,
  Romanov:2018eer}. Electrons are extracted from a photo-cathode in a
warm RF-gun immersed in a configurable longitudinal magnetic
field. After that, the beam is accelerated by two TESLA-type
superconducting cavities and one ILC-style cryomodule up to a kinetic
energy of 300~MeV. Transverse steering and focusing of the beam are
done with warm iron-based magnets. The FAST linac is equipped with a
wide variety of beam diagnostics, including electrostatic pickups, YAG
screens and optical transition radiation (OTR) foils.

During IOTA design and construction, special attention was given to
beam diagnostics. The main set consists of the following components:
\begin{itemize}
    \item wall-current monitors (WCMs)
    \item direct-current current transformers (DCCTs)
    \item 21 electrostatic pickups (BPMs)
    \item 2 synchrotron-radiation photomultipliers (PMTs)
    \item 8 synchrotron-radiation cameras
\end{itemize}
Precise beam position and shape measurements are necessary to tune
IOTA lattice parameters to the required
level~\cite{romanov2014lattice}. It turns out that the sensitivity of
the cameras is high enough to provide images even for a single
electron circulating in the ring for relatively short exposure times
(fractions of a second).

\begin{table}
  \centering
  \caption{IOTA electron beam parameters.}
    \begin{tabular}{|l|c|}
      \hline
      Parameter & Value \\
      \hline
      Perimeter & 39.96 m  \\
      Momentum & 100 MeV/$c$  \\
      Electron current & 0--4.8~mA  \\
      RF frequency & 30~MHz  \\
      RF voltage & 250~V  \\
      Betatron tunes, ($\nu_x, \nu_y$) & (0.28, 0.31)\\
      Synchrotron tune, $\nu_s$ & $3.5\cdot 10^{-4}$\\
      Damping times, ($\tau_x, \tau_y, \tau_s$) & (6.15, 2.38, 0.91)~s \\
      Horizontal  emittance (geom., RMS), $\epsilon_x$ & 36.6 nm \\
      Momentum spread (RMS), $\Delta p/p$    &   $8.4\cdot 10^{-5}$  \\
      Momentum compaction & 0.077 \\
      \hline
    \end{tabular}
  \label{tab:iota_params}
\end{table}

\begin{figure}
    \includegraphics[width=\columnwidth]{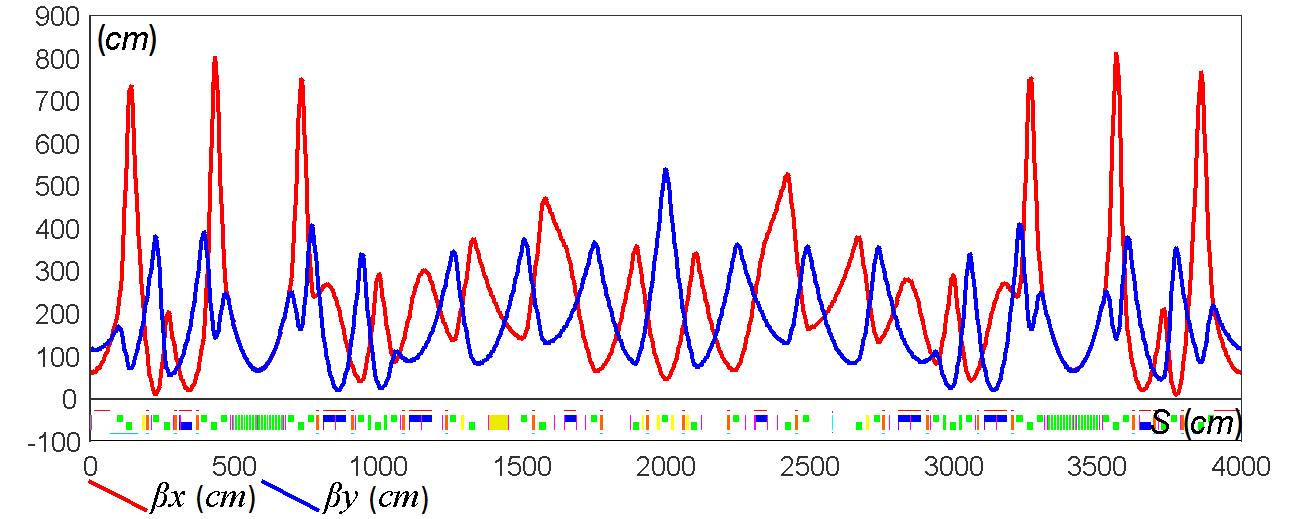}
    \caption{Horizontal (red) and vertical (blue) beta functions in
      the IOTA ring, as reconstructed with the LOCO method.}
    \label{fig:betaXY}
\end{figure}

\begin{figure}
    \includegraphics[width=\columnwidth]{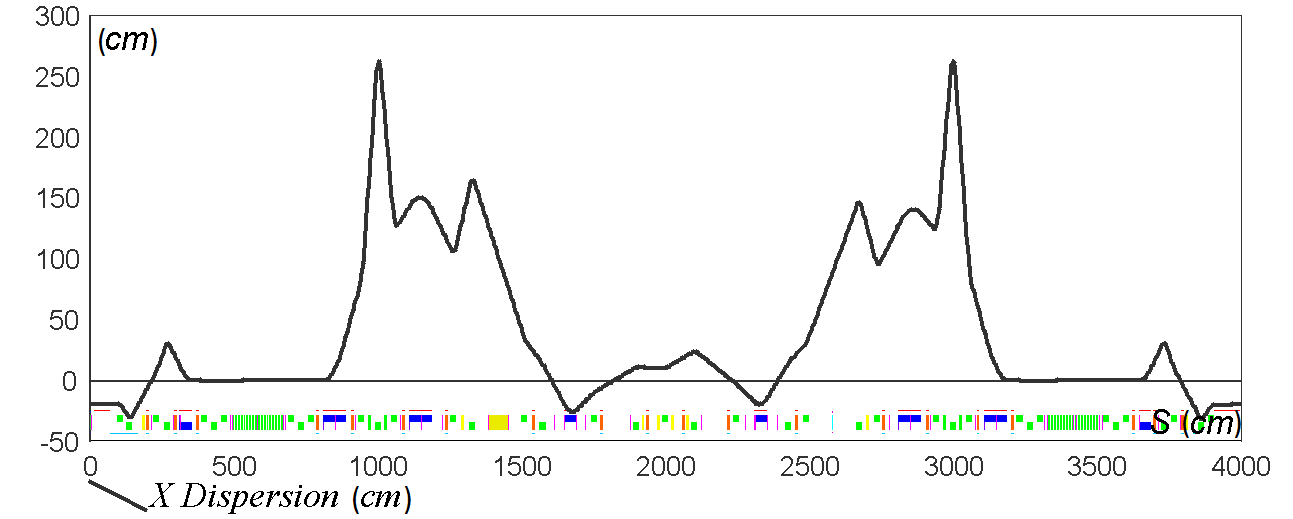}
    \caption{Horizontal dispersion in the IOTA ring, as reconstructed
      with the LOCO method.}
    \label{fig:dispX}
\end{figure}

During the experiments presented in this paper, IOTA operated with
100~MeV electrons and was configured for experiments with one special
nonlinear insertion, either a string of nonlinear Danilov-Nagaitsev
magnets~\cite{DN:PRSTAB, Valishev:Beams-doc-8871:2020} or an octupole
channel to generate a H\'enon-Heiles
potential~\cite{ValishevNagaitsevAntipov:Octupoles,
  kuklev2019Octupoles}. Table~\ref{tab:iota_params} lists the beam
parameters at the time of the experiments. Figure~\ref{fig:betaXY}
shows the horizontal and vertical beta functions and
Figure~\ref{fig:dispX} shows the horizontal dispersion. The lattice
parameters were analyzed and tuned using the LOCO technique
\cite{Safranek:1997mra,Sajaev:2005qw,romanov2017correction}. The
slight asymmetry in the lattice functions arises from the gradients in
the main dipoles, which were compensated by the quadrupoles.

\subsection{Synchrotron-Light Detection}

\begin{figure}
  \centering
  \includegraphics[width=\columnwidth]{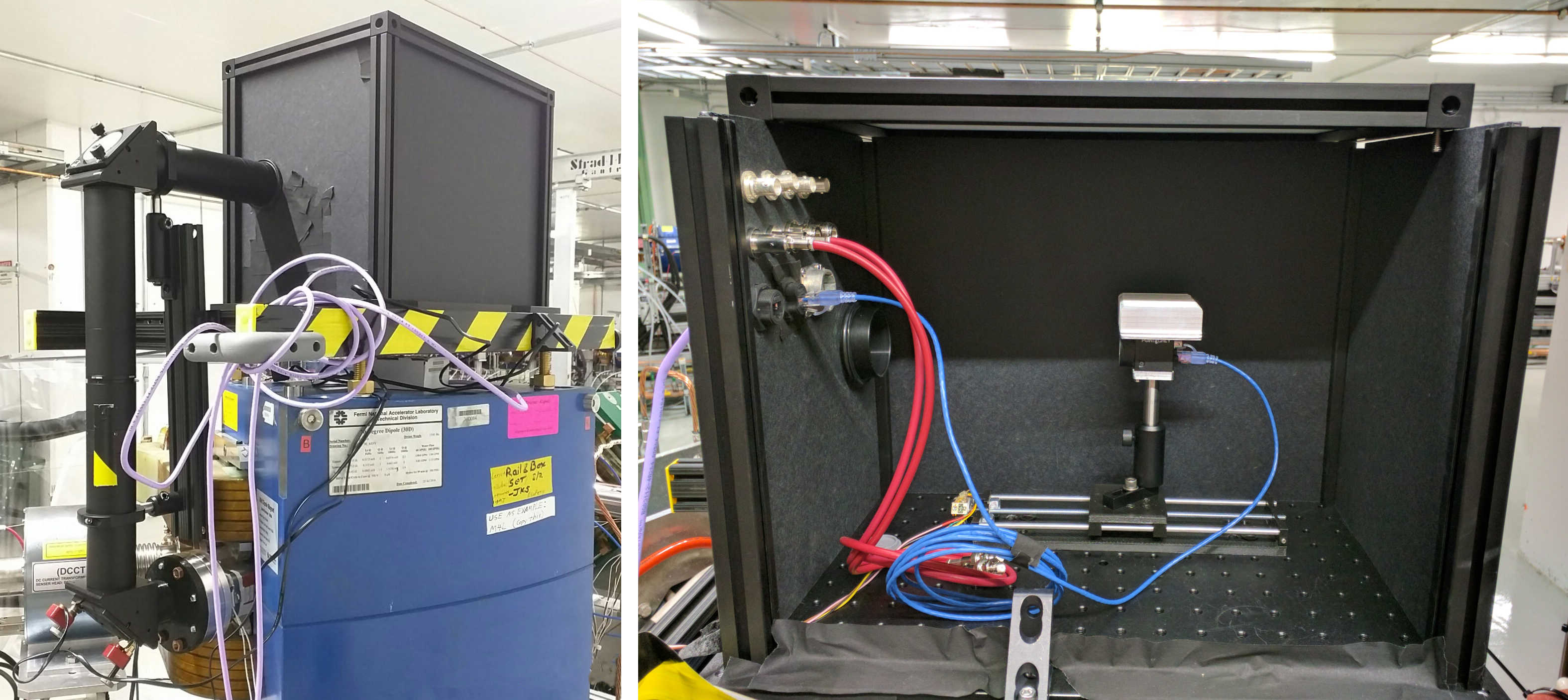}
  \caption {General view of a synchrotron-radiation station at a
    60-degree dipole (left). The interior of the dark box (right),
    with a passively cooled camera installed on a 3D-printed linear
    translation stage.}
  \label{fig:opticsAndCamera}
\end{figure}

Each of the 8~main dipoles in IOTA is equipped with synchrotron light
stations installed on top of the magnets themselves. The light out of
the dipoles is deflected upwards and back to the horizontal plane with
two 90-degree mirrors. After the second mirror, the light enters the
dark box, which is instrumented with customizable diagnostics, as
shown in Figure~\ref{fig:opticsAndCamera}. Custom 3D-printed viewport
adapters and 2-inch black tubes are used to connect optical components
and ensure light tightness. A focusing lens with a 40~cm focal length
is installed in the vertical lens tube that connects to the mirror
holders. At the time of the experiment, 7~magnets were equipped with
sensitive digital cameras. The main characteristics of the cameras are
listed in Table~\ref{tab:cameraParams}.

\begin{table}
  \centering
  \caption{Synchrotron-light camera parameters.}
    \begin{tabular}{|l|c|}
      \hline
      Parameter      & Value                  \\ 
      \hline
      Manufacturer   & Point Grey (now FLIR) \\
      Model          & BFLY-PGE-23S6M-C       \\
      Resolution     & 1920 $\times$ 1200 pixels \\
      Sensor         & Sony IMX249, CMOS      \\
      Pixel size     & \q{5.86}{\mu m}        \\
      ADC depth      & 12 bit                 \\
      Gain range     & 0 to 29.996~dB         \\
      Exposure range & \q{19}{\mu s} to 32~s \\
      Temporal dark noise (read noise)  & 7.11~$e^-$ \\
      Saturation capacity (well depth)  & 33106~$e^-$ \\
      Quantum efficiency at 500~nm & 83\% \\
      \hline
    \end{tabular}
  \label{tab:cameraParams}
\end{table}

A single 100~MeV electron circulating in the IOTA ring produces about
7\,500~detectable photons per second per main
dipole~\cite{Stancari:2017wox}, as confirmed by measurements with the
photomultipliers. This intensity is enough to be easily detected by a
PMT or by a sensitive digital camera. In IOTA, both PMTs and cameras
are routinely used to study beams with intensities from a single
electron up to the maximum current. About 10~photons per pixel are
necessary to exceed the background noise level on our cameras. This
puts a limit on the size of the light spot at the camera sensor.

\section{Amplitude Reconstruction Method}
\label{sec:reconstruction}

The oscillation amplitudes of the electron trajectories were obtained
by comparing synchronized sets of images with a model describing the
expected projections of these images onto the horizontal and vertical
axes.

The scaling factors for each camera were based on closed-orbit
responses of both cameras and electrostatic pickups to dipole trims
and RF frequency modulation, using the LOCO technique. A precise
modulation of the RF frequency allowed us to get an absolute
calibration from the dispersion measurements. This resolved
calibration degeneracies in the LOCO data set and gave absolute
calibrations for both BPMs and trims. 

There are several known effects that affect images quality. The first
group blurs an image and includes aberrations, diffraction, vignetting
and finite depth of field of the optics that images part of a curved
trajectory. There is also a possibility of nonlinear
distortions. After thorough alignment linearity and absence of
vignetting was verified with closed orbit scans, which is consistent
with relatively small camera sensors.

In the absence of linear coupling, the image of a single electron at a
position with beta functions $\beta_x$ and $\beta_y$ and dispersion
$D_x$ is the time average over the revolutions~$n$ of the
corresponding oscillations with mode-amplitudes~$A_x$, $A_y$, and
$A_p$:
\begin{equation}
\begin{split}
    x & = A_x \sqrt{\beta_x} \cos(\psi_{x,n}) + A_{\Delta p/p} D_x \cos(\psi_{p,n}),  \\
    y & = A_y \sqrt{\beta_y} \cos(\psi_{y,n}).  
\end{split}
\end{equation}
Therefore, the 1-dimensional probability density for a particle that
executes oscillations with amplitude~$R$ is:
\begin{equation}
  \rho_1(R, r) =    
  \begin{cases} 
    \frac{1}{\pi \sqrt{R^2-r^2}} & \mathrm{for}\ |r| \leq R, \\
    0 & \mathrm{for}\ |r| > R.
  \end{cases}
\end{equation}

In order to avoid confusion between mode amplitudes and oscillation
amplitudes corresponding to a specific beta-function or dispersion,
the latter will be denoted by capital letters of the corresponding
planes:
\begin{equation}
\begin{split}
    X_\beta&=A_x \sqrt{\beta_x},\\
    Y_\beta&=A_y \sqrt{\beta_y},\\
    X_p&=A_{\Delta p/p} D_x.   
\end{split}
\end{equation}
In the following discussion, a mode amplitude is assumed unless
indicated otherwise.

The image of a stable point-like
light source, or point spread function (PSF), was used to account
for the time-independent part of smearing. Because of the relatively
low signal to noise ratio, a normal distribution with a cutoff at
2~standard deviations~$L$ was used to model all blurring effects:
\begin{equation}
  \rho_{\mathrm{PSF}}(L, r) =
  \begin{cases}
    K \exp{\left(-\frac{r^2}{2 L^2}\right)} & \mathrm{for}\ |r| \leq 2 L, \\
    0 & \mathrm{for}\ |r| > 2 L.
  \end{cases}
\end{equation}
Here $K$ is a coefficient that normalizes the integral of the
PSF function to~1.

The resulting model distribution for one-mode oscillations is:
\begin{equation}
  \rho_{1\mathrm{PSF}}(R, L, r) =
  \int \rho_1(R, \tilde{r}) \, \rho_{\mathrm{PSF}}(L, \tilde{r}-r) \, d\tilde{r}.
\end{equation}

If the particle undergoes two independent oscillations with
frequencies that are not in a resonance, such as during combined
synchrotron and betatron oscillations, the convolutions of two one-mode
densities and a PSF has to be calculated:
\begin{equation}
 \rho_{2\mathrm{PSF}}(R_1, R_2, L, r)  
  = \int \left[ \int  \rho_1(R_1, l) \rho_1(R_2, l-\tilde{r}) dl
  \right] \, \rho_{\mathrm{PSF}}(L, \tilde{r}-r) \, d\tilde{r}.
\end{equation}

\begin{figure*}
  \centering
  \begin{subfigure}[b]{0.48\textwidth}
    \centering
    \includegraphics[width=\textwidth]{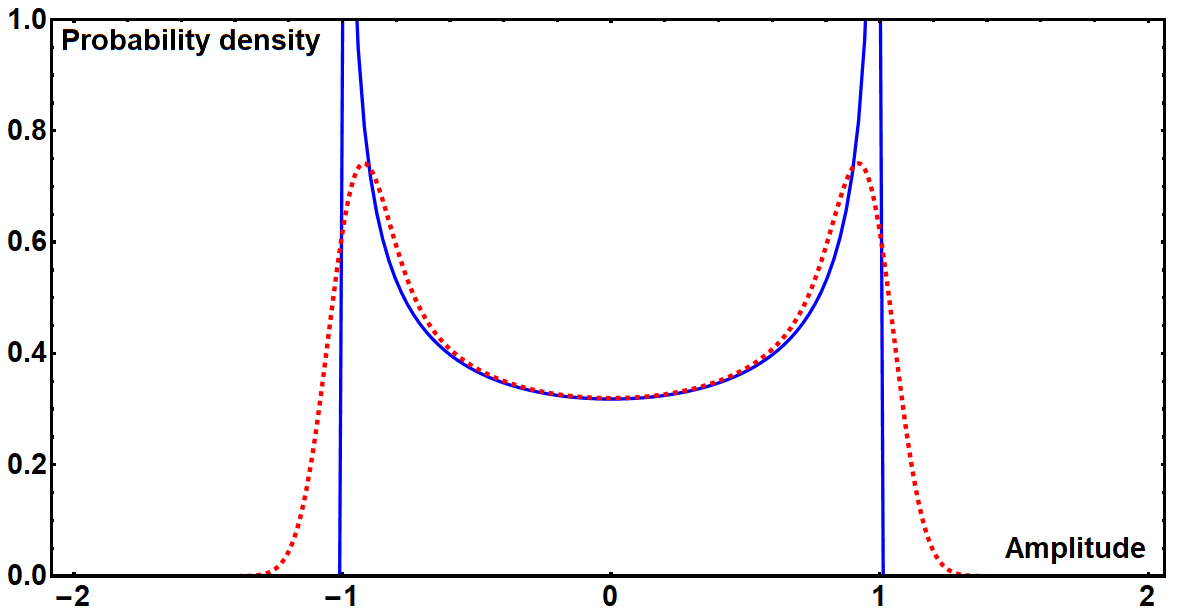}
    \caption{One oscillation mode}    
  \end{subfigure}
  \quad
  \begin{subfigure}[b]{0.48\textwidth}  
    \centering 
    \includegraphics[width=\textwidth]{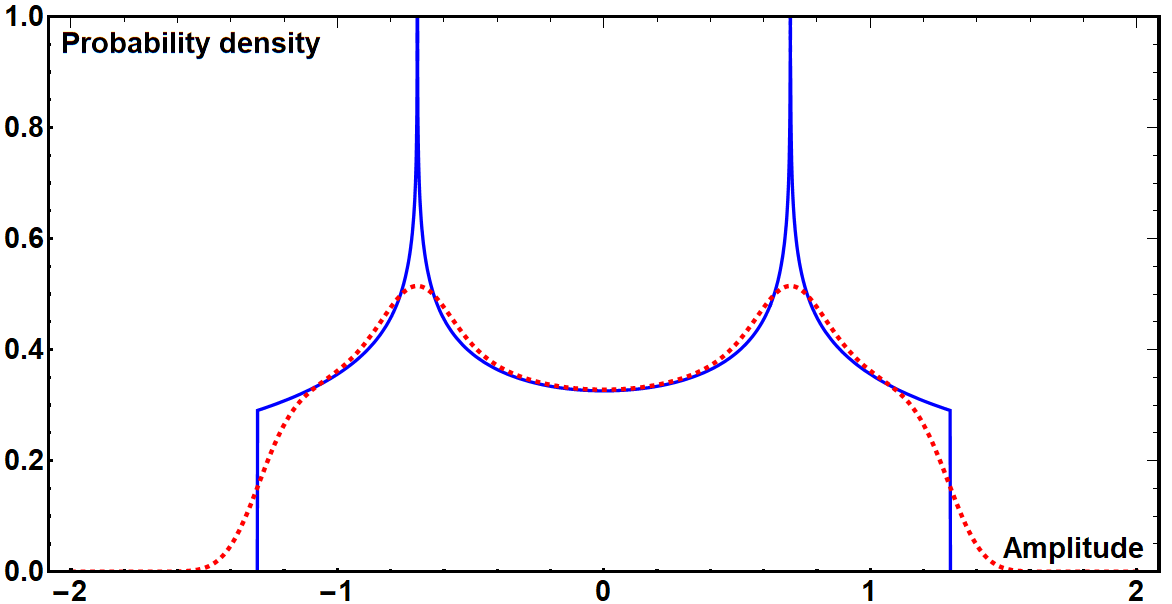}
    \caption{Two oscillation modes}
  \end{subfigure}
  \caption{Model projections for the cases of one and two modes, with
    and without smearing due to a Gaussian PSF (dashed red lines and
    solid blue lines, respectively). The amplitude of the first mode
    is 1 (arb.\ units), the second amplitude is 0.3, and the standard
    deviation of the PSF is 0.1.}
  \label{fig:modelProjections}
\end{figure*}

Figure~\ref{fig:modelProjections} shows model projections for
$\rho_{1}$, $\rho_{1\mathrm{PSF}}$, $\rho_{2}$ and
$\rho_{2\mathrm{PSF}}$. The projection of two independent oscillations
has a distinct shape, with peaks at $\pm(R_1-R_2)$ and shoulders that
extend to $\pm(R_1+R_2)$ in the case of a narrow PSF. Having a feature
of the image projections that depends linearly on a small amplitude
(the momentum spread contribution) makes it easier to resolve both
amplitudes, especially in comparison with the case of a beam profile
with a nearly Gaussian shape. In the latter case, the smooth profile
depends on the small amplitude only quadratically. In the absence of
characteristic features, a good absolute calibration of the cameras is
necessary to resolve the size contribution due to momentum spread.

To model the horizontal and vertical projections from $N$ cameras,
$3+5N$ parameters were used:
\begin{itemize}
\item the amplitudes $A_x$, $A_y$ and $A_p$ and
\item 5 camera-specific parameters,
  \begin{itemize}
  \item $I$, the common signal intensity for both planes
  \item $L_{x}$ and $L_y$, the standard deviations of the PSF functions
  \item $x_0$ and $y_0$, the closed orbit offsets.
  \end{itemize}
\end{itemize}

\section{Statistical Properties of Single-Particle Dynamics}
\label{sec:statistics}

A beam formed by linear focusing forces and frequent small kicks
due to synchrotron-radiation damping acquires a normal distribution in
position. The 1-dimensional projection of such a distribution is also
normal and can be represented as follows:
\begin{equation}
  \rho(r) = \frac{1}{\sqrt{2\pi \sigma_r^2}}
  \exp{\left(-\frac{r^2}{2 \sigma_r^2}\right)}.
  \label{eq:gaussDensity}
\end{equation}

In the case of horizontal and vertical projections with negligible
coupling and vertical dispersion, coordinate variances due to betatron
and synchrotron oscillations can be calculated as follows:
\begin{equation}
\begin{split}
  \sigma_x^2 &= \sigma_{x,\beta}^2+\sigma_{x,D}^2=\epsilon_x \beta_x +
    (D_x \sigma_{\Delta p/p})^2, \\
  \sigma_y^2 &= \sigma_{y,\beta}^2=\epsilon_y \beta_y.
  \label{eq:gaussSigmas}
\end{split}
\end{equation}

Each particle in such a beam has 3~modes of oscillation. For each
mode, in the case of a normal distribution in phase space, the
probability density~$p$ of mode-amplitude~$A$ (square root of the
Courant-Snyder invariant) can be expressed as a function of the
equilibrium mode-amplitude~$A_0$ as follows:
\begin{equation}
  p(A) = \frac{A}{A_0^2} \exp{\left(-\frac{A^2}{2A_0^2}\right)}.
  \label{eq:gaussAmplitudeDensity}
\end{equation}

Beam emittance, standard deviation of coordinates and equilibrium amplitude are
related according to these relations:
\begin{equation}
  \begin{split}
    \epsilon &= A_0^2 \\
    \sigma &= \sqrt{\epsilon \beta} = A_0\sqrt{\beta}.
  \end{split}
  \label{eq:emittanceSigmaAndAmplitude}
\end{equation}

The first, second and fourth moments of the mode-amplitude distribution
(Eq.~\ref{eq:gaussAmplitudeDensity}) will be used in later analyses:
\begin{equation}
  \left\langle A \right\rangle = \sqrt{\frac{\pi}{2}} A_0 , \quad
  \left\langle A^2 \right\rangle = 2 A_0^2 , \quad
  \left\langle A^4 \right\rangle = 8 A_0^4.
\label{eq:amplitudesMoments}
\end{equation}

For each degree of freedom, the damping time~$\tau$ is closely related
to the amplitude autocorrelation function. The latter can be directly
calculated from each amplitude time series, resulting in a damping
time estimation that does not require a complex analysis of individual
events.

The correlation between squared amplitudes~$A_1^2$ at time~$t_1$ and
$A_2^2$ at time $t_2 = t_1+ \Delta t$ is
\begin{equation}
  \langle A_1^2 A_2^2 \rangle_{t_1, t_2} =
  \langle A^2(t_1) A^2(t_1+\Delta t) \rangle_{t_1,\Delta t},
\end{equation}
where the averaging is done over initial amplitudes at time~$t_1$ and
their changes after a time interval~$\Delta t$.

On average, a squared amplitude converges to the equilibrium
value $\left\langle A^2 \right\rangle=2 A_0^2$, independently of its
initial value. Therefore, its average at time~$t_2$ can be expressed
as
\begin{equation}
  \langle A_2^2 \rangle_{t_2} = 2  A_0^2 + (A_1^2-2 A_0^2) e^{-2\Delta t/\tau} .
\end{equation}

Averaging over time~$t_1$ yields
\begin{equation}
\begin{split}
  \langle A_1^2 A_2^2 \rangle_{t_1,t_2} &= \langle A_1^2
  \left[2 A_0^2 + (A_1^2-2 A_0^2) e^{-2\Delta t/\tau}\right] \rangle_{t_1} \\
  & = \left( \langle A_1^4\rangle -2 \langle A_1^2\rangle A_0^2 \right)
  e^{-2\Delta t/\tau} +2 \langle A_1^2\rangle A_0^2 
   = 4 A_0^4 \left( 1+\exp^{-2\Delta t/\tau} \right).
\end{split}
\label{eq:autocorrelation}
\end{equation}

Expression~\ref{eq:autocorrelation} is used in
Section~\ref{sec:results} to obtain estimates of the damping times. In
these experiments, the measured amplitudes are averaged over the
camera exposure times. Additionally, there is a finite resolution
limit for both maximum and minimum detectable amplitudes. These
constraints have a noticeable effect especially in the vertical plane
because, in the case of decoupled lattice, the vertical equilibrium
emittance is very small.

\section{Experimental Results}
\label{sec:results}

\subsection{Single Electron Injection}

Several steps were taken to inject a small number of electrons in
IOTA. First, the laser of the FAST linac photo-injector was switched
off, so that only dark current was generated. Three OTR foils were
inserted along the injector to further reduce intensity. Fine control
of the attenuation was achieved by using the last quadrupole before
the IOTA injection kicker. As a result, it was possible to obtain a
distribution for the number of injected particles with peak and
average near 1~electron. If needed, it was also possible to remove
electrons from the circulating beam, one at a time, by carefully
lowering the voltage of the IOTA RF cavity and then quickly restoring
it to the nominal value.

\begin{figure}
  \centering
  \begin{subfigure}{0.49\textwidth}
    \includegraphics[width=\textwidth]{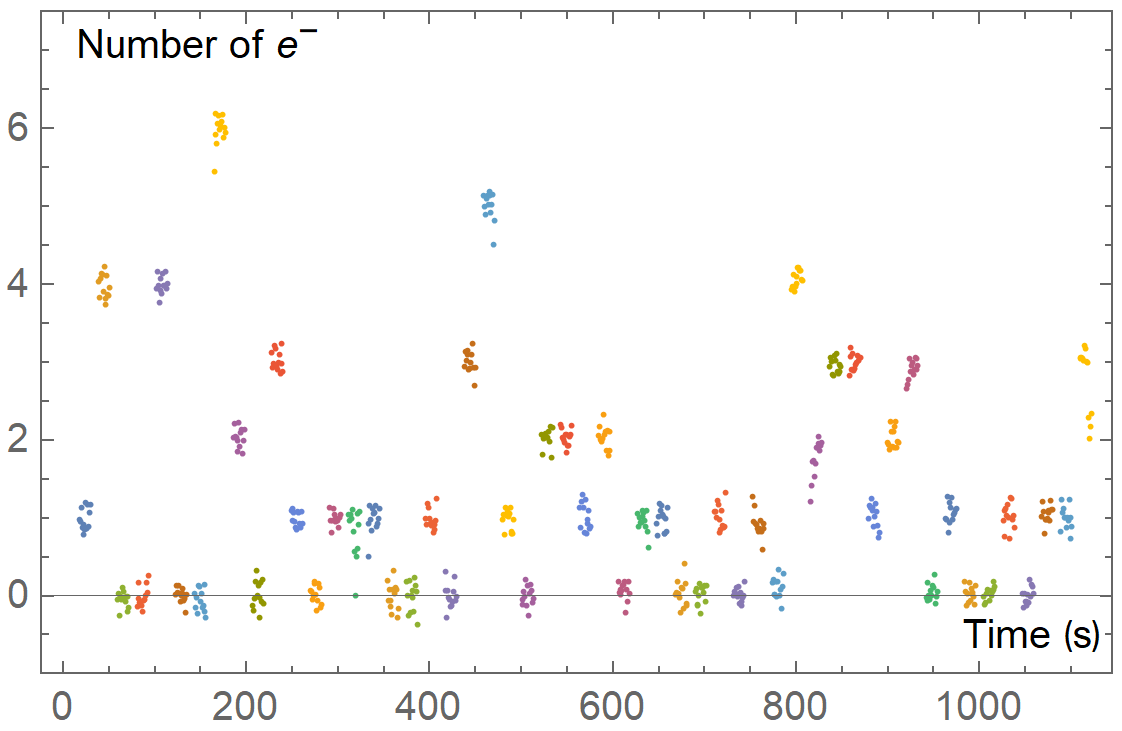}
    \caption {Integrated pixel intensity from the region of interest
      of one of the IOTA cameras, normalized to the one-electron
      intensity. Points from different injections are indicated by
      different colors.}
    \label{fig:1e_injection_time}
  \end{subfigure}
  \hfill
  \begin{subfigure}{0.49\textwidth}
    \includegraphics[width=\textwidth]{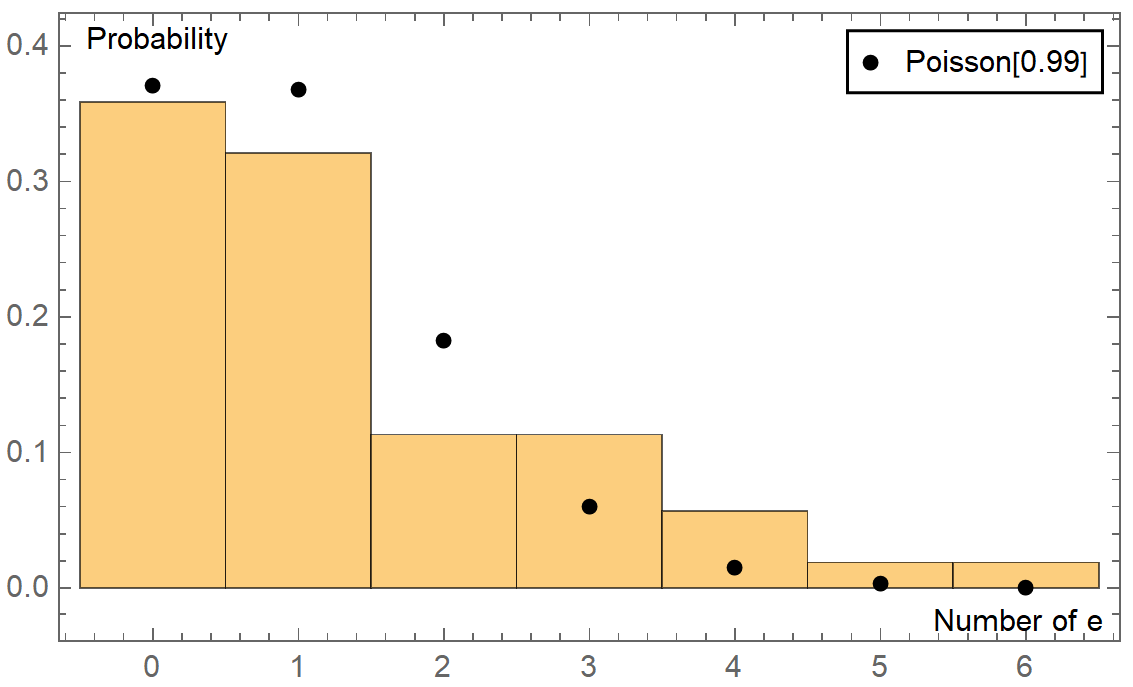}
    \caption {Distribution of the observed number of injected
      electrons in IOTA (yellow histogram). For comparison, the
      best-fit Poisson distribution is also shown (black dots).}
    \label{fig:1e_injection_probability}
  \end{subfigure}  
  \caption{Injections of a few electrons in IOTA.}
  \label{fig:1e_injection}
\end{figure}

To study the distribution of the number of injected electrons, a
series of 53~injections with a period of 21~s were made. The
integrated intensity in the region of interest (ROI) of one camera
sensor was used to measure the number of injected electrons. The
period of the injections was chosen to allow the transverse
oscillations to damp, so that the beam images would fit well within
the ROI.  Figure~\ref{fig:1e_injection_time} shows the camera signal
as function of time for different injections. Images were taken at a
rate of about 1~Hz, with exposure times of
1~s. Figure~\ref{fig:1e_injection_probability} shows the measured
distribution of the number of captured electrons. The measured
probability of single-electron injections was~32\%, which is close to
the maximum probability of~36.8\% for an ideal Poisson process.

\subsection{Beam Lifetime and Camera Calibration with a Known
  Countable Number of Circulating Electrons}

The lifetime of a beam with about one hundred electrons was measured
during the same shift as the single-electron studies. The camera
settings were the same. The maximum number of electrons in the ring
for this data set was chosen to avoid saturation of the core
pixels. At the end of the measurement, the beam was dumped by reducing
the voltage of the RF cavity to measure the cameras background levels.

\begin{figure*}
  \includegraphics[width=\textwidth]{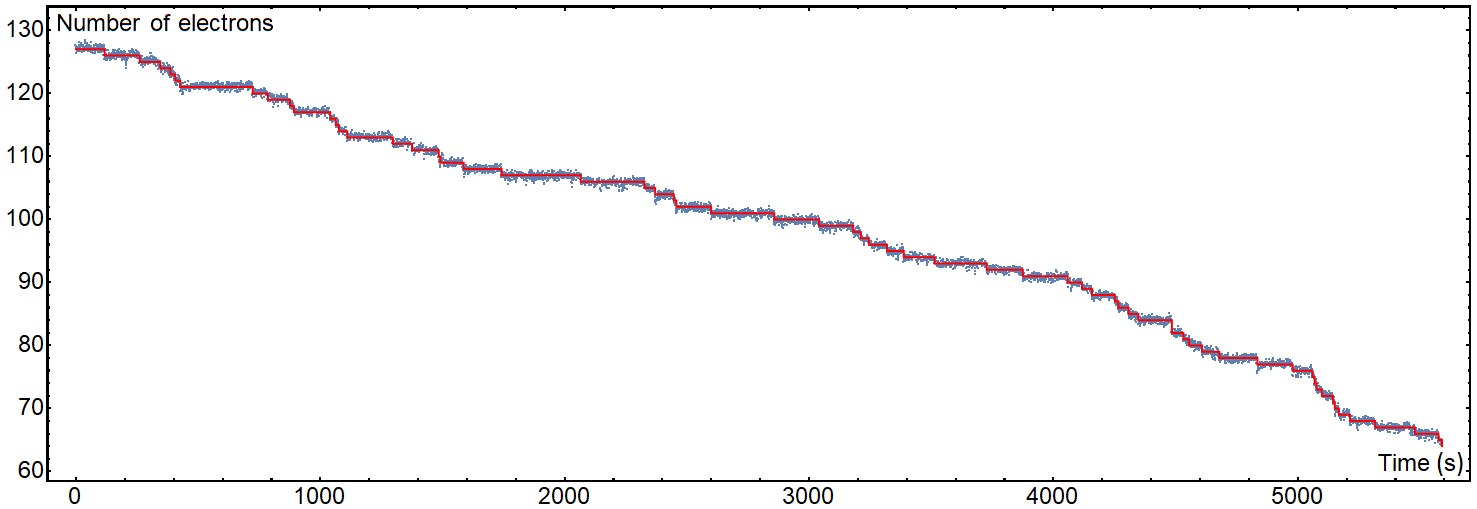}
  \caption {Number of circulating electrons in IOTA as a function of
    time. Beam intensity measured as the sum of pixel signals in the
    regions of interest of all 7~cameras, normalized to the
    single-electron step size. The red line is a best fit to the data
    with integer numbers of electrons.}
  \label{fig:lifetime}
\end{figure*}

Figure~\ref{fig:lifetime} shows the total signal from all cameras in
an elliptical region of interest (10 and 12.5~beam standard deviations
in horizontal and vertical, respectively), normalized to the size of
the discrete steps due to single-electron losses. Hot pixels were
replaced with neighbor medians and a global median filter with radius
of 2~pixels was applied.

\begin{figure}
  \begin{subfigure}{\columnwidth}
    \includegraphics[width=\linewidth]{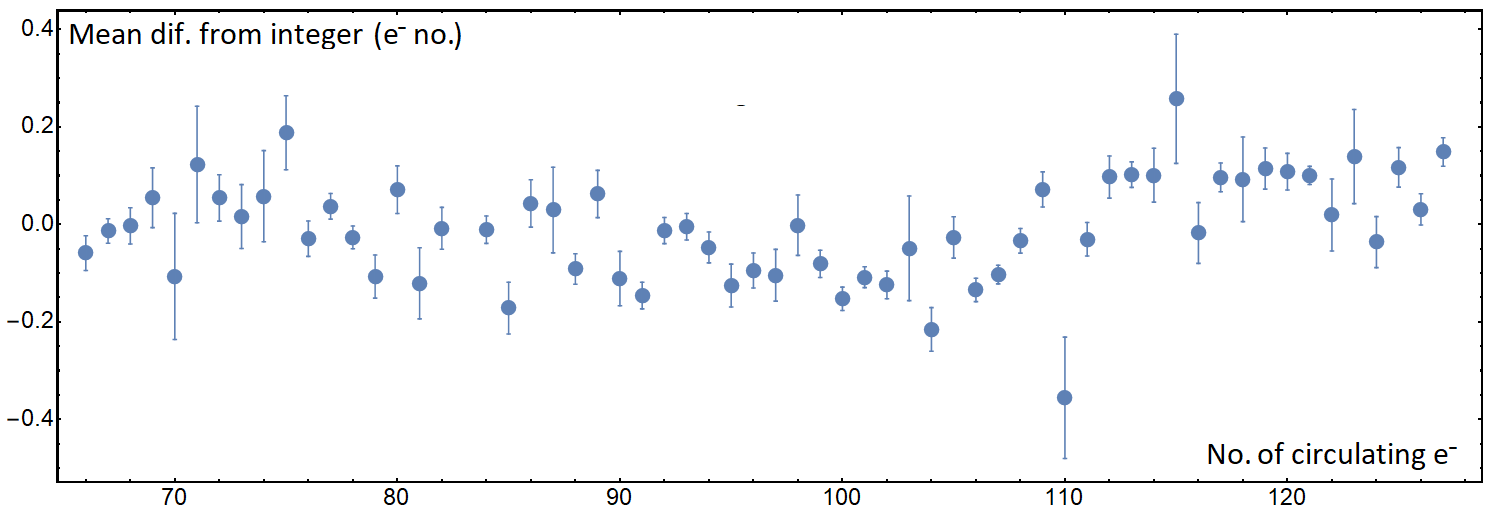}
    \caption{Deviation of the calibrated signal from an integer number
      of electrons vs.\ number of electrons.}
    \label{fig:linearityCalibration}
  \end{subfigure}
  \begin{subfigure}{\columnwidth}
    \includegraphics[width=\linewidth]{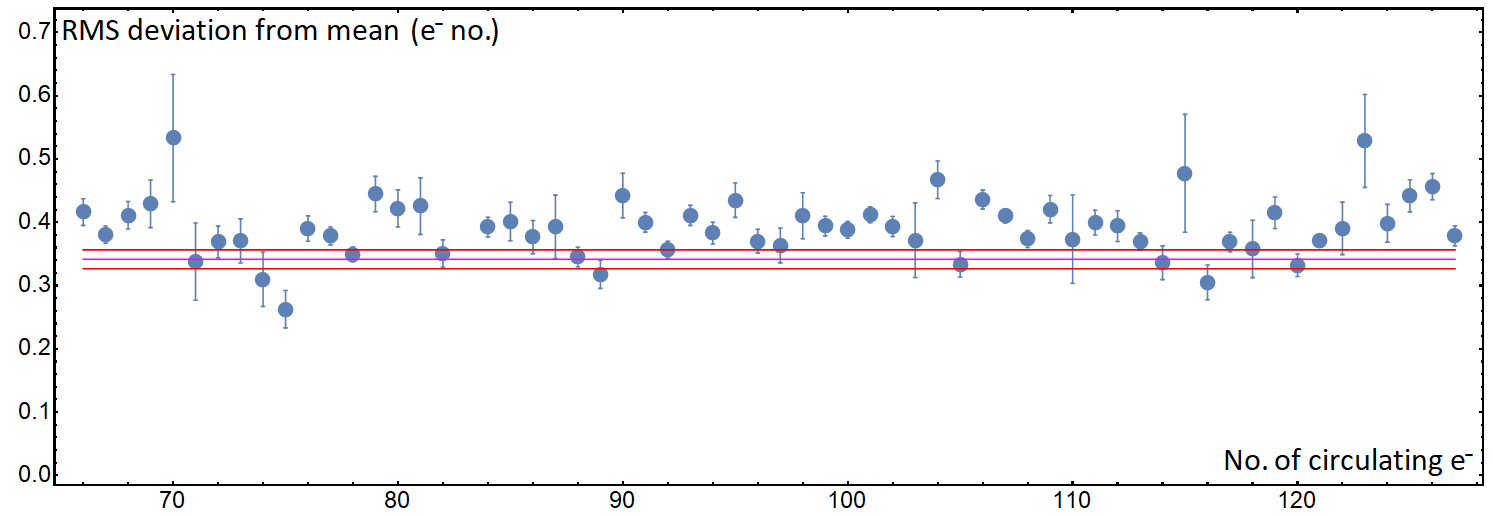}
    \caption{Fluctuations of the calibrated signal vs.\ number of
      electrons. The red horizontal lines represent the known baseline
      noise level without signal and its uncertainty.}
    \label{fig:snRatio}
  \end{subfigure}
  \caption{Linearity and noise of the absolute intensity calibration
    of the camera signals using known numbers of circulating
    electrons.}
\end{figure}

Because of the discrete nature of single-electron losses, the beam
intensity evolution does not follow the usual exponential decay
predicted by the continuous model. At these low intensities, a
maximum-likelihood estimate of beam lifetime can be calculated from
the time intervals between individual
losses~\cite{Stancari:FN-lifetime:2020}. The relative statistical
uncertainty is equal to the inverse square root of the number of
observed steps. For this data set, the beam lifetime was 9100(1200)~s
or 2.52(32)~h.

Because of the discrete steps in intensity, the data set can be
subdivided into periods with constant integer numbers of
electrons. From these subsets, an absolute calibration of the optical
system vs.\ circulating beam current can be performed.  Linearity was
verified by observing the deviations of the calibrated mean signal
from the corresponding expected integer values
(Figure~\ref{fig:linearityCalibration}). Figure~\ref{fig:snRatio}
shows the standard deviation of the fluctuations of the measured beam
intensity in each subset. The dependence on the number of electrons
was rather weak, which implies that the contribution of photon
statistics was small and that signal fluctuations were dominated by
readout noise. For future studies, several methods to improve the
signal-to-noise ratio are being considered, such as reduction of the
beam spot size on the sensors, cooling of the sensors themselves, and
upgraded cameras with better matrices and electronics.

\subsection{Evolution of Single Electron Oscillation Amplitudes}

\begin{figure*}
  \includegraphics[width=0.95\textwidth]{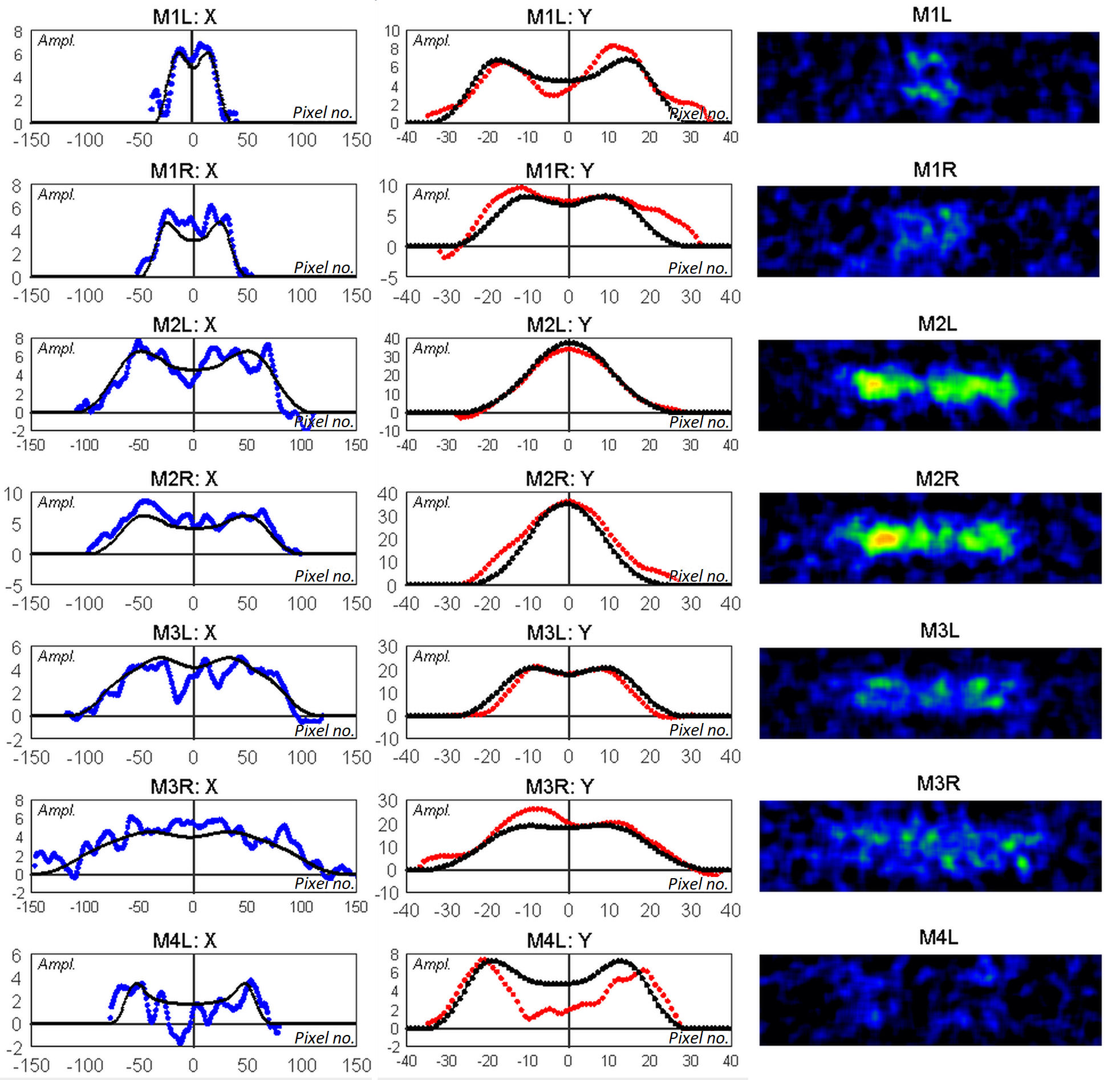}
  \caption {A set of synchronized images from 7~cameras (right
    column), together with their horizontal (left column) and vertical
    (central column) projections. The blue and red curves show the
    experimental data, whereas the black curves represent the fitted
    model projections. All images were cropped to the size of 300 by
    80~pixels.}
  \label{fig:sampleSet}
\end{figure*}

\begin{figure*}
  \begin{tabular}{c}
    \includegraphics[width=\textwidth]{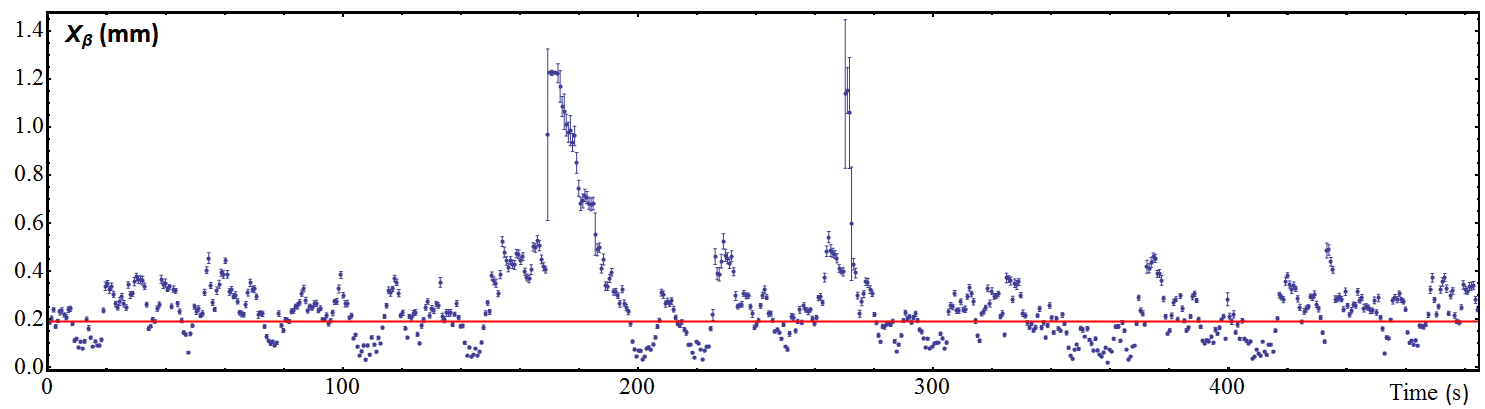} \\
    \includegraphics[width=\textwidth]{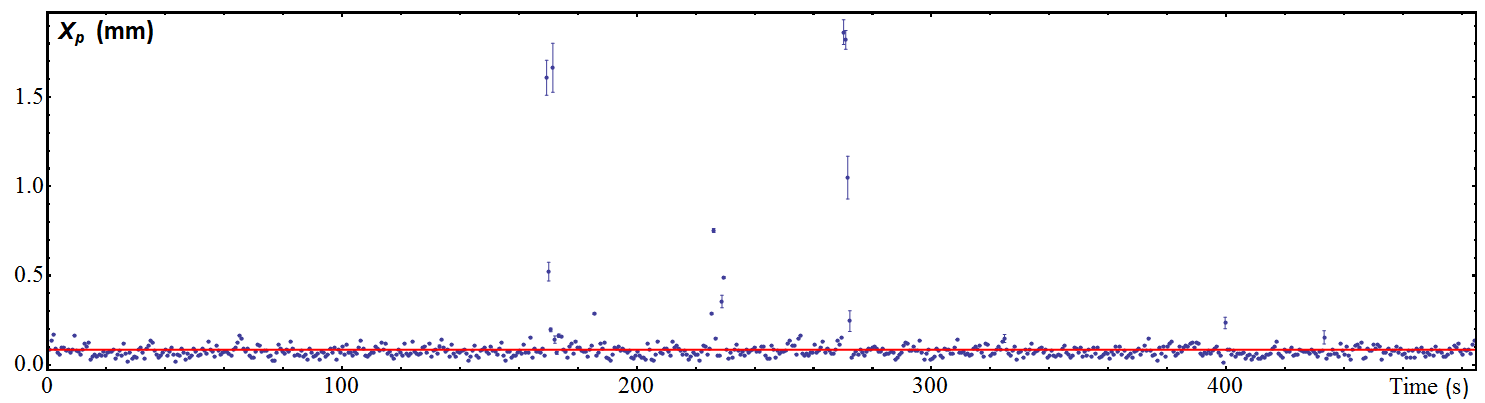} \\
    \includegraphics[width=\textwidth]{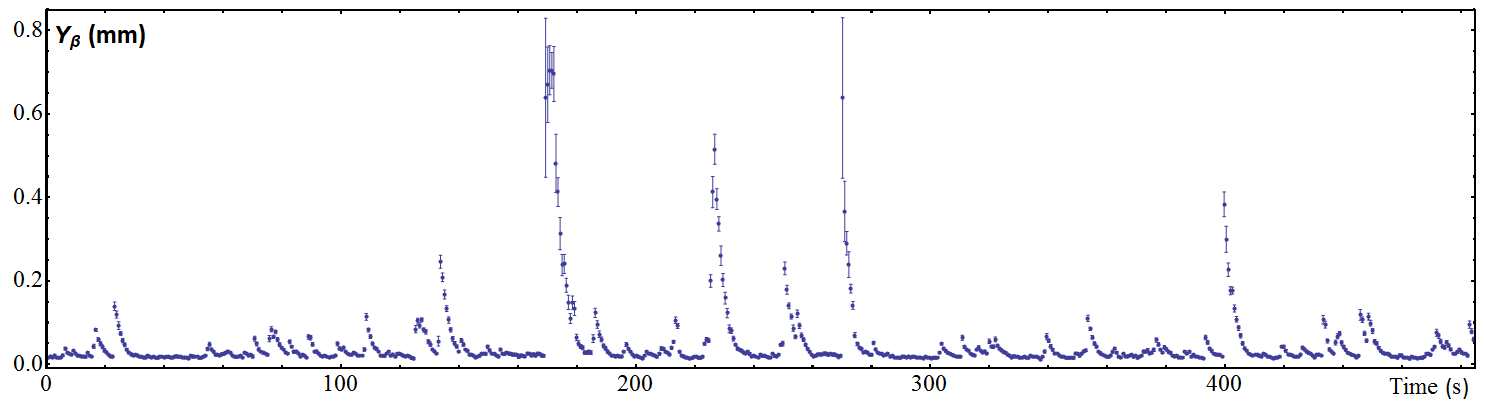}
  \end{tabular}
  \caption{Evolution of the measured oscillation amplitudes for a
    single electron in IOTA. The top and bottom plots show the
    horizontal and vertical betatron amplitudes at a location with 1~m
    beta functions. The middle plot shows the contribution to
    horizontal amplitude from synchrotron oscillations, at a location
    with 1~m dispersion. The red lines show the calculated equilibrium
    amplitudes for 100 MeV electrons, accounting only for fluctuations
    of synchrotron radiation.}
  \label{fig:amplTimeEvolution}
\end{figure*}

The analysis presented here is based on a series of 2876~sets of
images. Each set consists of synchronized images from 7~cameras with
exposures of 0.5~s and delays of 0.2~s between them. The delay was
necessary to read out and save raw frames to a hard drive. A median
filter with a 2~pixel radius and a moving average filter with a radius
of 5~pixels were applied to each image to reduce noise. To stabilize
the fitting algorithm, the sizes of the point-spread functions and the
total intensities at each camera were determined from a limited subset
of images with intermediate values of all 3~amplitudes. To improve
convergence of the fitting algorithm, bounded parameter ranges were
used. The lower limit of amplitude was set well below the actual
resolution power. The upper limit was determined by the sensitivity of
the cameras and was selected at the level where the electron signal
became small compared to the background noise. The upper limits on the
closed orbit offsets were set at a level that exceeded realistic
closed orbit jitter by an order of magnitude and that was triggered
only in cases when the electron was excited beyond the detection
limit.

For robust estimates of the uncertainties on the model parameters, we
used the bootstrap method. The method is also useful to detect
anomalies in the fitting process. For each set of synchronized image
projections, 25~synthetic bootstrap samples were generated and the
corresponding distribution of fit parameters was calculated.

As an example, Figure~\ref{fig:sampleSet} shows a synchronized set of
images from the 7~cameras, together with their horizontal and vertical
measured and fitted projections.

Figure~\ref{fig:amplTimeEvolution} shows the fitted amplitudes and
their statistical uncertainties during the first 8~min of
observations. The 3~planes exhibit different patterns, in general. The
horizontal plane has a damping time that is large compared to the
camera exposure time. Amplitudes are large enough to be reliably
resolved. Random fluctuations, mostly driven by fluctuations of
synchrotron radiation emission, can be seen. Synchrotron oscillations
have the shortest damping time, which is comparable to the exposure
duration. Therefore, the reconstructed amplitude is close to the
equilibrium value. In the vertical plane, amplitude excitations from
synchrotron radiation are very small --- what is observed are
relatively sparse interactions with the residual gas.

A quantitative analysis is presented below.

\subsubsection{Horizontal Betatron Oscillations}

\begin{figure}[t]
  \includegraphics[width=\columnwidth]{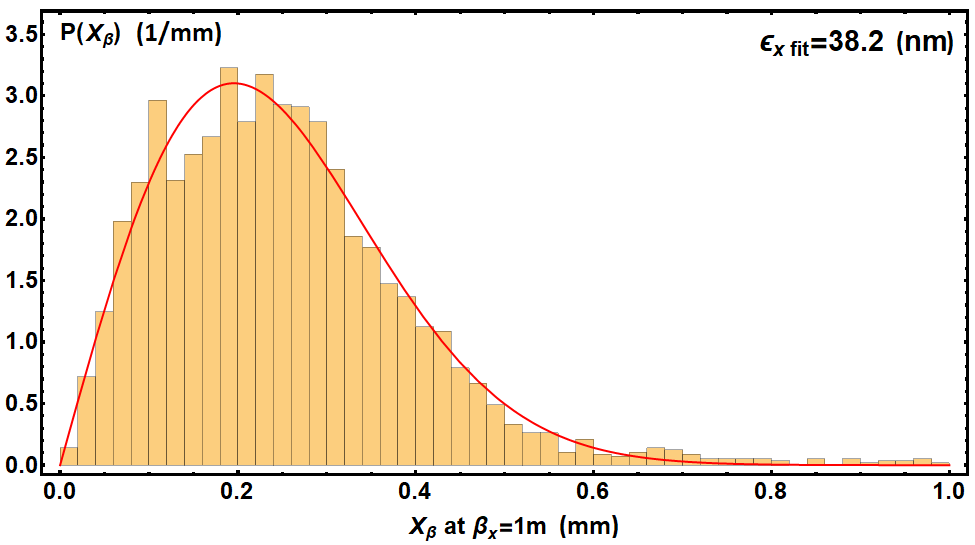}
  \caption{Histogram of measured horizontal betatron amplitudes at a
    location with 1~m beta function. The red curve is the best fit to
    a model with normal distributions in phase space, yielding an
    emittance of 38.2~nm.}
  \label{fig:xAmplHistogram}
\end{figure}

\begin{figure}
  \includegraphics[width=\columnwidth]{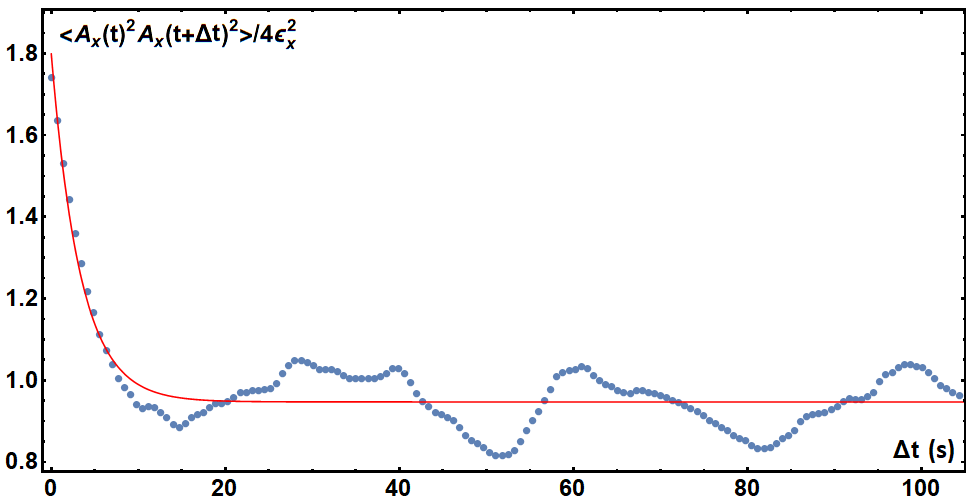}
  \caption{Autocorrelation of squared horizontal amplitudes vs.\
    delay.}
  \label{fig:xAmplCorrelation}
\end{figure}

The damping time in the horizontal plane was much longer than the
exposure duration. Moreover, the average amplitude was well above the
resolution of the cameras. Most of the kicks from the residual gas
were not resolvable on top of a large horizontal emittance determined
by the random kicks from synchrotron radiation photons.

The amplitude time series also contains 3~cases of excitations to very
large amplitudes (for example, at the time mark of about 175~s in
Figure~\ref{fig:amplTimeEvolution}). It took about 20~s to return to
normal dynamics, with a time constant about 2~times longer than the
synchrotron damping time. This suggests that these events were the
result of glitches in IOTA components.

Figure~\ref{fig:xAmplHistogram} shows a histogram of all horizontal
amplitudes, together with the best fit of the amplitude probability
density for a Gaussian distribution in phase space
(Eq.~\ref{eq:gaussAmplitudeDensity}). The resulting equilibrium
emittance is $\epsilon_x = \q{38.2}{nm}$. This value is very close to
the lattice model prediction (Table~\ref{tab:iota_params}), confirming
that horizontal dynamics was dominated by synchrotron radiation. This
slightly larger horizontal emittance could be explained by lattice
imperfections and heating from collisions with the residual gas. A
higher beam energy might also have this effect, but this explanation
was ruled out by further analysis (see also below).

Figure~\ref{fig:xAmplCorrelation} shows the autocorrelation of
squared oscillation amplitudes in the horizontal plane as a function
of delay. The resulting damping time, according to the model described
in Section~\ref{sec:statistics}, is $\tau_x = \q{6.7(0.6)}{s}$. This
value is in good agreement with the lattice model
(Table~\ref{tab:iota_params}).

\subsubsection{Synchrotron  Oscillations}

\begin{figure}[t]
  \includegraphics[width=\columnwidth]{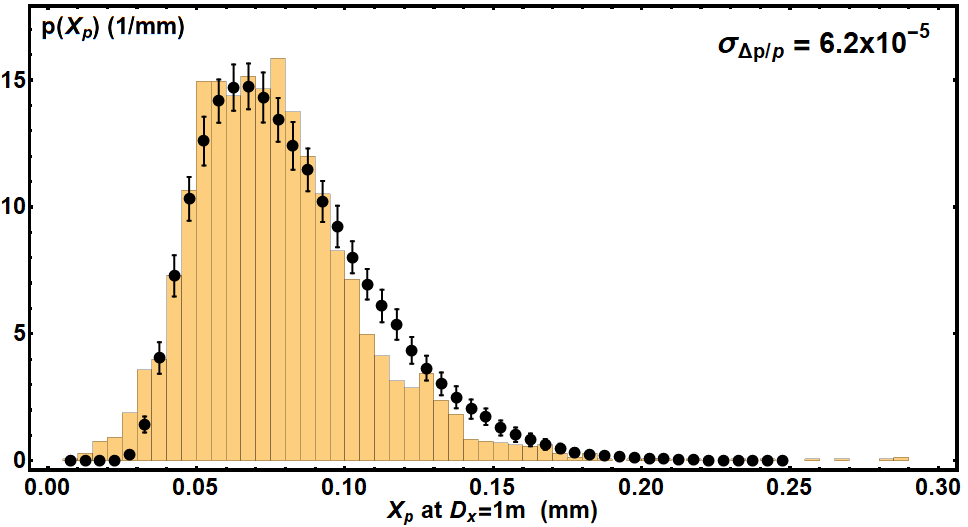}
  \caption{Histogram for measured amplitudes of synchrotron
    oscillations projected on the horizontal plane at a location with
    1~m dispersion. The the black dots show expected distribution
    obtained from a simulation of evolution of synchrotron amplitude,
    averaged over exposure time; error bars show standard deviation
    for the number of measured amplitudes equal to that in the
    experimental data set.
    $\sigma_{\Delta p/p} = 6.2 \cdot 10^{-5}$.}
  \label{fig:pAmplHistogram}
\end{figure}

\begin{figure}
  \includegraphics[width=\columnwidth]{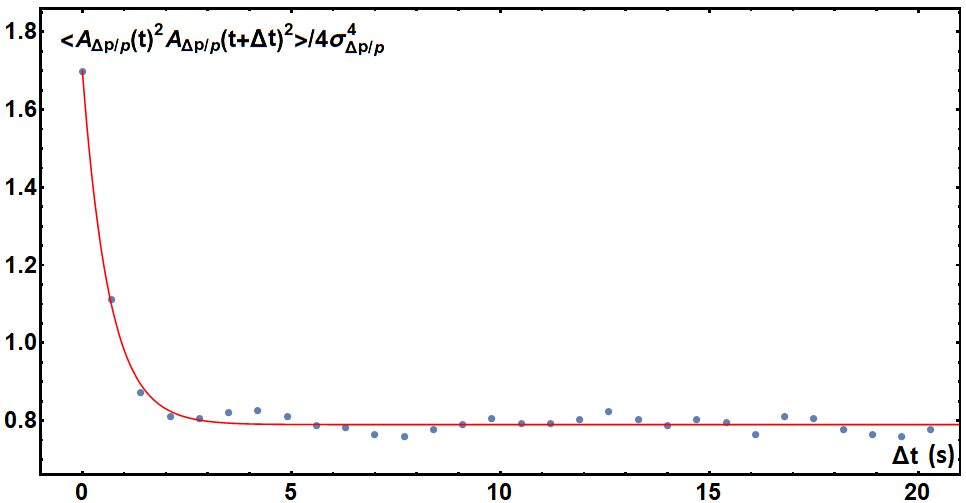}
  \caption{Autocorrelation of squared synchrotron amplitudes vs.\
    delay.}
  \label{fig:pAmplCorrelation}
\end{figure}

A histogram of reconstructed amplitudes for synchrotron oscillations
is presented in Figure~\ref{fig:pAmplHistogram}. Since the exposure
duration of 0.5~s was comparable to the damping time of energy
oscillations (about 1~s), the reconstructed values are averaged
amplitudes and they tend to be close to the equilibrium value of
$\left\langle A_{\Delta p/p} \right\rangle$. To compare the observed
distribution with the expected one, a simple Monte~Carlo simulation of
synchrotron amplitude evolution was done over a time span that is 100
times longer than the extent covered by the analyzed experimental
data. Simulations were done for the same damping time and mean
amplitude values as those extracted from the experimental data. The
simulation was used to plot both the probability distribution of the
synchrotron amplitude and the standard deviation corresponding to the
finite size of the experimental data set.

The equilibrium momentum spread was evaluated as follows:
\begin{equation}
  \sigma_{\Delta p/p} = \left\langle A_{\Delta p/p} \right\rangle
  \sqrt{\frac{2}{\pi}} = 6.2 \cdot 10^{-5}.
\end{equation}

The autocorrelation function for squared amplitudes is shown in
Figure~\ref{fig:pAmplCorrelation}. A fit to the model function
(Eq.~\ref{eq:autocorrelation}) gives an amplitude damping time of
$\tau_{\Delta p/p} = \q{1.28(6)}{s}$.

\subsubsection{Vertical Betatron Oscillations}

The calculated vertical amplitude equilibrium values and fluctuations
due to synchrotron radiation were negligible compared to camera
resolutions, because coupling was strongly suppressed and tunes were
set away from the coupling resonance. The observed amplitude evolution
in the vertical plane was dominated by scattering on the residual gas.

\begin{figure}
  \includegraphics[width=\columnwidth]{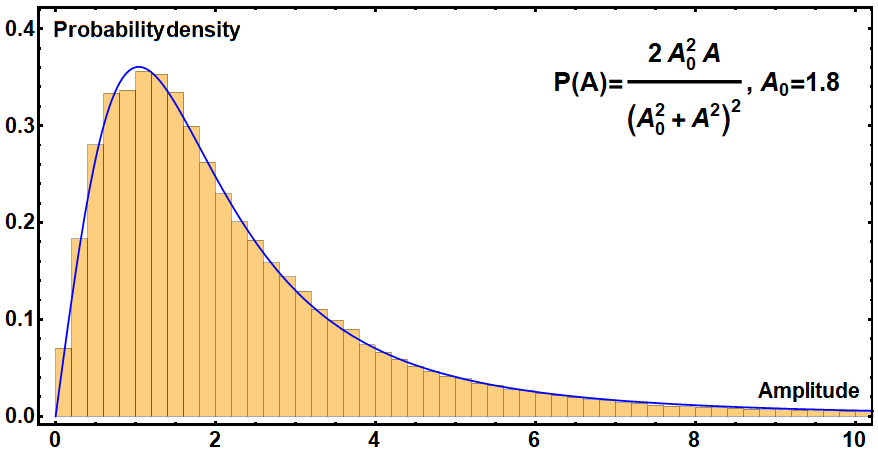}
  \caption{Monte~Carlo simulation of the distribution of amplitudes
    due to gas scattering. The solid blue line shows the best fit
    using the empirical formula of Eq.~\ref{eq:gasAmplitudeDensity}.}
  \label{fig:gasScatterModel}
\end{figure}

The probability of scattering to an amplitude greater than $A_0$ is
proportional to $1/A^2_0$. For the case when a particle has enough
time, on average, to damp to low amplitudes before experiencing the
next large scattering event, the probability to observe an amplitude
smaller than the kick amplitude is proportional to $1/A$. Therefore,
the tail of the distribution formed by relatively sparse residual gas
collisions is proportional to $1/A^3$. At the same time, for small
amplitudes, the distribution should be proportional to $A$ because of
small but nonzero coupling.  The amplitude probability density should
therefore be described by an empirical formula such as
\begin{equation}
  p(A) = \frac{2 A_0^2 A}{(A_0^2+A^2)^2}.
  \label{eq:gasAmplitudeDensity}
\end{equation}
A simple Monte~Carlo simulation of an electron interacting with the
residual gas, for parameters close to those in IOTA, was used to
verify that the empirical formula describes quite well the
distribution of amplitudes. The results are shown
in~Figure~\ref{fig:gasScatterModel}.

\begin{figure}
  \includegraphics[width=\columnwidth]{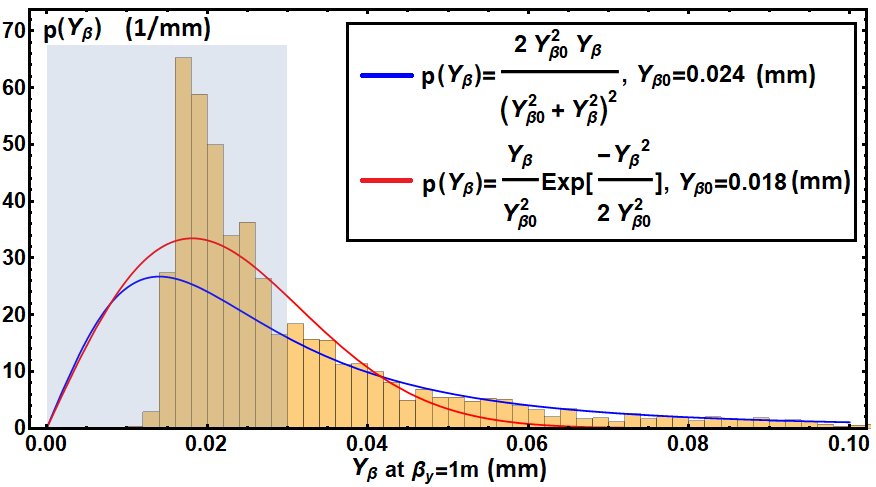}
  \caption{Measured distribution of vertical betatron amplitudes
    (yellow histogram), normalized to a location with a 1~m beta
    function. The blue curve is the best fit to the empirical formula
    (Eq.~\ref{eq:gasAmplitudeDensity}), while the red curve represents
    the best fit to a Gaussian beam
    (Eq.~\ref{eq:gaussAmplitudeDensity}). The shaded area was excluded
    from the fits because of the limited camera resolution (see
    text).}
  \label{fig:yAmplHistogram}
\end{figure}

Figure~\ref{fig:yAmplHistogram} shows a histogram of the measured
vertical amplitude distribution, normalized to a location with
$\beta_y = \q{1}{m}$. The finite resolution of the optical systems
allowed us to reliably resolve amplitudes above \q{30}{\mu m}
(relative to the same $\beta_y = \q{1}{m}$). Bins corresponding to
amplitudes greater than \q{30}{\mu m} were used to obtain the best fit
to the empirical distribution of Eq.~\ref{eq:gasAmplitudeDensity}. For
comparison, the same bins were also used to get a best fit of
amplitude distributions for a Gaussian beam
(Eq.~\ref{eq:gaussAmplitudeDensity}).

\begin{figure}
  \includegraphics[width=\columnwidth]{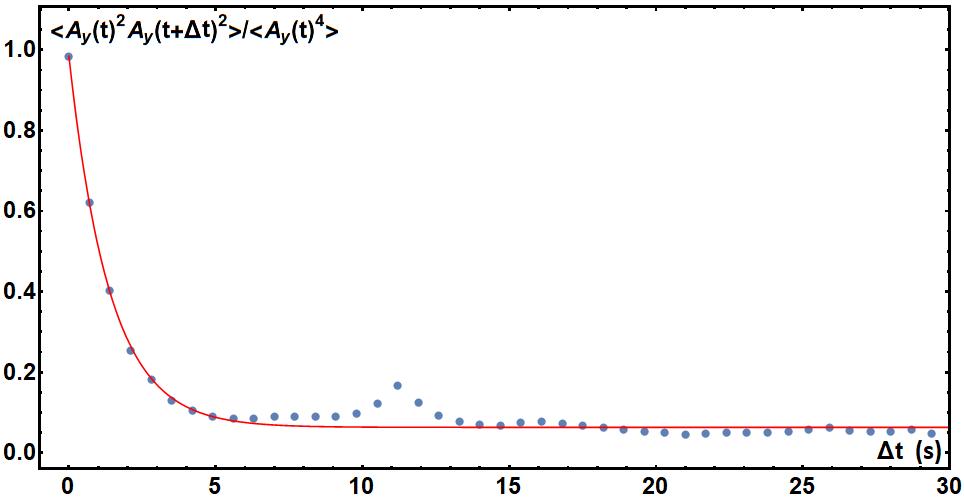}
  \caption{Autocorrelation of squared vertical amplitudes.}
  \label{fig:yAmplCorrelation}
\end{figure}

Figure~\ref{fig:yAmplCorrelation} shows the autocorrelation of squared
vertical amplitudes. For the empirical distribution of
Eq.~\ref{eq:gasAmplitudeDensity}, all moments greater than the first
are infinite. Therefore, mathematically, Eq.~\ref{eq:autocorrelation}
is not applicable. In practice, in an experiment or simulation, all
moments are finite due to limited observation time or finite aperture
and the relation
$\langle A_1^4\rangle \gg 2 \langle A_1^2 \rangle A_0^2$ will hold. In
the case of negligible background, the time constant of the
exponential decay is unchanged. From this data set, we obtained the
vertical amplitude damping time $\tau_y = \q{2.77(10)}{s}$.

\begin{figure}
\centering
\begin{subfigure}[t]{.46\textwidth}
  \centering
  \includegraphics[width=\columnwidth]{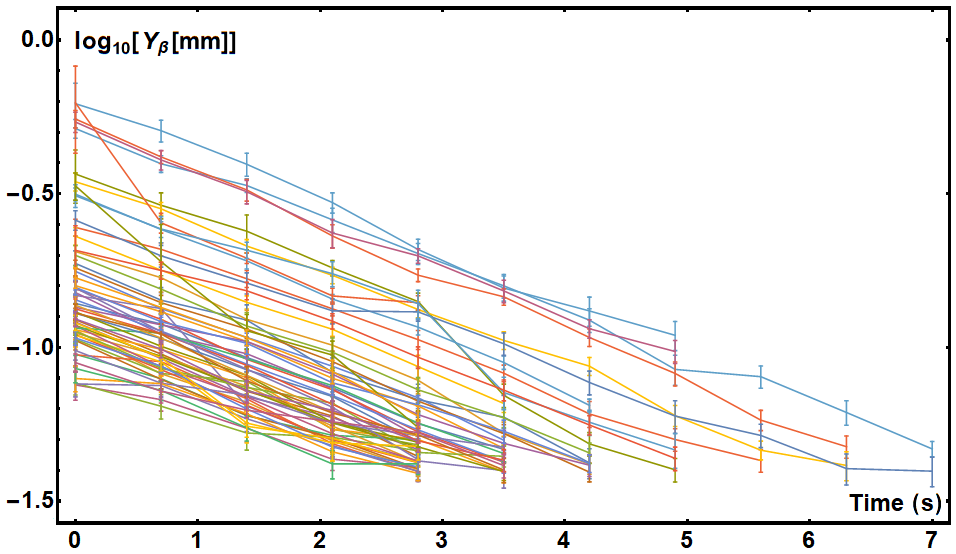}
  \caption{Vertical amplitude damping for 59 large-angle scattering
    events on the residual gas.}
  \label{fig:yAmplScatterDecay}
\end{subfigure}
\hfill
\begin{subfigure}[t]{.51\textwidth}
  \centering
  \includegraphics[width=\columnwidth]{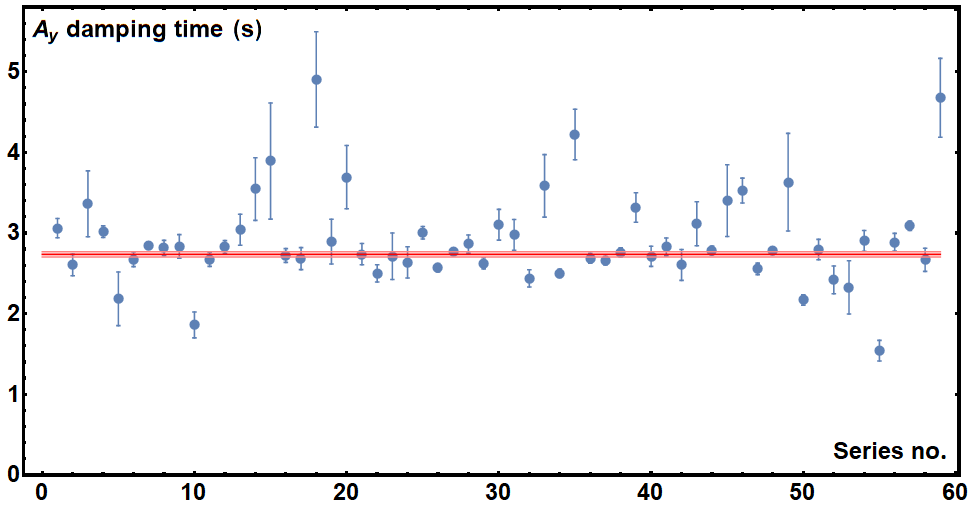}
  \caption{Damping times for all 59 series of scattering events. The
    average value and uncertainty are represented by the horizontal red and pink lines.}
  \label{fig:yAmplDampTimes}
\end{subfigure}
\caption{Analysis of vertical damping times after scattering events.}
\label{fig:yAmpl}
\end{figure}

Another way to extract the vertical damping time is a direct analysis
of the amplitude decays after large-amplitude scattering on the
residual gas. The full data set contained 59~series that satisfied the
following criteria:
\begin{itemize}
    \item $A_y^2$ was between 1.5~nm and 400~nm;
    \item $A_y^2$ decreased compared to the previous measurement;
    \item there were at least 5~points in the series.
\end{itemize}
Figure~\ref{fig:yAmplScatterDecay} shows all 59~selected series of
amplitudes normalized to a 1~m beta function. The corresponding
damping times, assuming zero vertical equilibrium emittance, are
presented in Figure~\ref{fig:yAmplDampTimes}. The resulting vertical
amplitude damping time is $\tau_y = \q{2.730(32)}{s}$, which agrees
with the results obtained from the autocorrelation analysis.

One can use the strong dependence of the vertical damping time on beam
energy to estimate the energy of the electrons circulating in IOTA.
Strongly suppressed transverse coupling makes the vertical damping time
almost independent of the lattice configuration. For a flat ring with
revolution period~$T_0$, the vertical damping rate is
\begin{equation}
  \frac{1}{\tau_y} =
  \frac{1}{2} \frac{C_\gamma E^3}{2 \pi T_0} \oint\frac{ds}{r^2},
  \label{eq:yDampTimeFromEnergy}
\end{equation}
with
$C_\gamma = (4\pi/3) r_e / (m_e c^2)^3 = \q{8.85 \times 10^{-5}}{m /
  GeV^3}$.  The steep energy dependence gives favorable scaling for
error propagation from damping time to energy. The two main systematic
errors are transverse coupling and the actual magnetic field geometry
of the main dipoles, which affects the integral of the squared
curvature of the closed orbit. The latter effect was investigated
using magnetic measurements and models of fringe fields of the main
dipoles. Integration of 3D magnetic fields gives a 7.6\% correction
compared to the hard-edge model. For both types of main dipoles, a
comparison between the measured and modeled fringe fields, accounting
for known trims, yields a correction of less than 1\% to the
calculated energy value. To address the systematics due to coupling,
we note that the horizontal damping time was longer than the vertical
one. Therefore coupling between horizontal and vertical planes tends
to increase the vertical damping time and therefore reduce the value
of the beam energy. Experimental data showed that the vertical
emittance contribution from coupling did not exceed 0.2~nm. If one
introduces coupling randomly, so as to create a 0.2~nm increase in
vertical emittance, the corresponding vertical damping time increases,
on average, by 0.8\%. The resulting IOTA beam energy estimate is
$E = \q{97.90 \pm 0.40 \mathrm{(stat)} \pm 0.35 \mathrm{(syst)}}{MeV}$.

\subsection{Residual Gas Characteristics and Machine Aperture}

Information about residual gas pressure and composition can be
extracted from beam lifetime and from the statistics of small-angle
scattering events in the vertical plane. Both processes are dominated
by Coulomb scattering on residual gas nuclei. Other effects, such as
bremsstrahlung and ionization losses, have significantly smaller cross
sections.

The key difference between small amplitude scattering seen in the
vertical plane and large amplitude kicks leading to particle loss is
the impact parameter. The impact parameter determines the amount of
nuclear screening by atomic electrons. Characteristic effects of
nuclear screening and the relatively large number of small-angle
impacts lead to a rough estimate of the gas composition. Anomalies in
the gas composition may indicate leaks or other problems with the
vacuum system. On the other hand, scattering to intermediate and large
angles, when screening effects are small and the electron is lost,
allow one to estimate the available aperture, which is helpful to
understand the machine performance. Of course, the relatively low
number of intermediate- and large-angle collisions results in large
uncertainties on the aperture estimates.

The differential cross section for single Coulomb scattering with
screening is given by the following approximate expression (see for
instance Ref.~\cite{chao1999handbook}, Section~3.3.1):
\begin{equation}
  \frac{d\sigma}{d\Omega} \simeq \frac{4 Z^2 r_e^2}{\gamma^2\beta^4}
  \frac{1}{(\theta_x^2+\theta_y^2+\thscreen^2)^2},
  \label{Eq:sigma}
\end{equation}
where $\thscreen \simeq \alpha Z^{1/3} / (\beta \gamma)$.

For large dynamic apertures $\thxymax \gg \thscreen$ and relatively
small threshold vertical angles $\theta_{y0} \ll \thxymax$,
integration over all horizontal angles $\theta_x$ and integration over
vertical angles $\theta_y$ exceeding $\theta_{y0}$ gives the following
cross section:
\begin{equation}
  \sigma(Z,\theta_{y0}) \simeq 
  \frac{4 \pi Z^2 r_e^2}{\gamma^2\beta^4\thscreen^2}
  \left( 1 - \frac{\theta_{y0}}{\sqrt{\theta_{y0}^2+\thscreen^2}} \right).
\end{equation}

The frequency of excitations to an amplitude  exceeding
$A_y$ can be calculated as follows, taking into account the
beta functions and partial pressures around the ring:
\begin{equation}
  \frac{1}{\tau(A_y)}
  = \frac{c \beta}{\Pi} \oint_{\mathrm{ring}}
  \frac{ \neff(s) \cdot \sum_Z  k_Z(s) \cdot
    \sigma\left(Z, \frac{A_y}{\sqrt{\beta_y(s)}}\right) }{\sum_Z k_Z(s)
    \cdot Z^2} \, ds.
\end{equation}
Here $\Pi$ is the perimeter of the ring;
$\neff (s) = \sum_Z n(Z, s) \cdot Z^2$ is the effective gas density
along the ring; $k_Z(s)$ are un-normalized relative gas
densities. (Multiplying and dividing by $\sum_Z k_Z(s) \cdot Z^2$
allows one to use arbitrary nonzero coefficients as relative
densities. For example, it is convenient to set one of the
coefficients $k_Z(s_0)$ to 1.)

\begin{figure}
  \centering
  \enspace\includegraphics[width=0.9\textwidth]{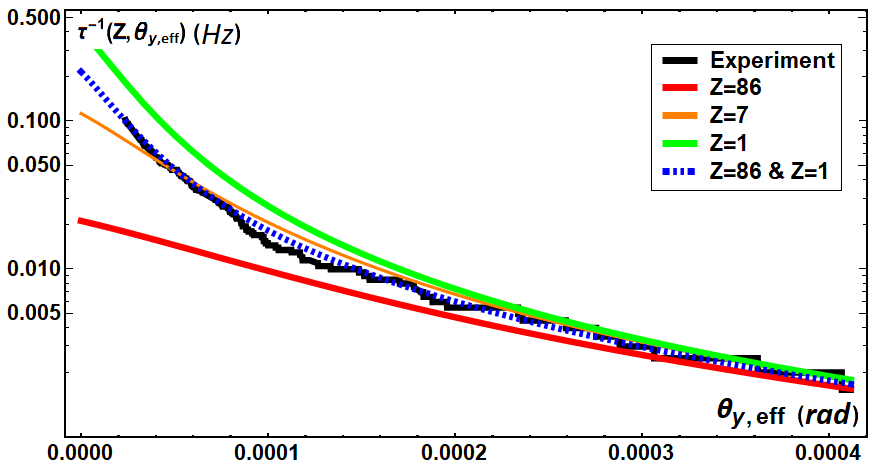}
  \caption{Measured and calculated dependence of scattering
    frequencies on the effective vertical angle~\thyeff.}
  \label{fig:excitationFrequencies}
\end{figure}

\begin{figure}
  \includegraphics[width=0.95\textwidth]{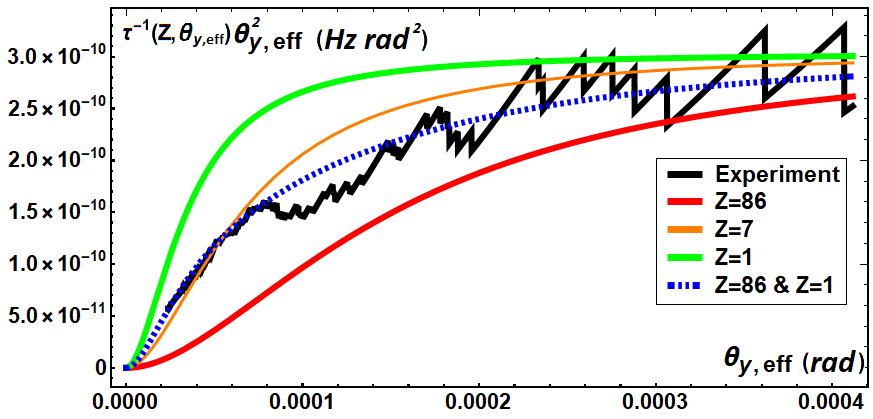}
  \caption{Dependence on the effective vertical angle~\thyeff\ of the
    scattering frequencies multiplied by~$\thyeff^2$.}
  \label{fig:excitationFrequenciesNorm}
\end{figure}
  
Experimentally, the dependence of the excitation frequencies on
amplitude was extracted from the vertical amplitude time series,
partially shown in Figure~\ref{fig:amplTimeEvolution}. First, the
whole data array was filtered to exclude periods of a few very large
excitations that made it impossible to detect small amplitude
changes. Then, each amplitude jump was recorded. Special attention was
given to properly treat transition frames, in which the electron
oscillated with two significantly different amplitudes. Such frames
typically result in a reconstructed amplitude that is in-between
initial and excited values. A simplistic detection algorithm might
count them as two events with smaller kicks. Instead, an interpolation
method was used to get an estimate of the actual scattering amplitude.

Figure \ref{fig:excitationFrequencies} shows the dependence of
scattering frequencies on the effective vertical angle
$\thyeff = A_{y}/ \sqrt{\left\langle\beta_y\right\rangle}$,
assuming constant partial pressures around the IOTA
ring. Figure~\ref{fig:excitationFrequenciesNorm} shows the same plot
with normalized frequencies (i.e., frequencies multiplied
by~$\thyeff^2$). Without screening effects, the normalized frequencies
should be independent of scattering amplitude. The small number of
large-amplitude kicks results in noticeable discrete steps in the
experimental data.

The plots show the experimental scattering frequencies together with
4~model curves for atomic numbers~$Z$ of 1~(hydrogen), 7~(nitrogen),
86~(radon) and for a combination of~$Z$ values~1 and~86. This
combination, with partial coefficients $k_1 = 1$ and
$k_{86} = 1.3(4) \times 10^{-4}$, showed the best fit, especially in
the small-angle region, where most of the scattering events are
concentrated. The resulting effective residual gas density is
$\neff = \q{7.0(1.6) \times 10^8}{cm^{-3}}$.

The ring aperture can be estimated from the large-amplitude scattering
cross section, together with the known beam lifetime and effective
residual gas density. For an elliptic aperture that requires kick
angles much larger than~\thscreen\ for a particle to be lost,
integration of Eq.~\ref{Eq:sigma} and averaging around the ring gives
the following cross section:
\begin{equation}
  \tilde{\sigma}(Z, \Axmax, \Aymax) \simeq 
  \frac{2 \pi Z^2 r_e^2}{\gamma^2\beta^4}
  \left\langle
    \frac{\beta_x(s)}{\AAxmax} + \frac{\beta_y(s)}{\AAymax}
  \right\rangle
\end{equation}
The available data does not allow one to distinguish between restrictions
in the vertical and the horizontal planes. The measured lifetime of
9100~s corresponds to apertures
$\AAxmax = \AAymax \simeq \q{12}{\mu m}$, assuming equal maximum
amplitudes in both planes.

\section{Conclusions}
\label{sec:conclusions}

For the first time, to our knowledge, the dynamics of a single
electron in a storage ring was tracked in all 3~dimensions using
high-resolution synchrotron-light images acquired with digital
cameras.

Data was taken at the Fermilab Integrable Optics Test Accelerator
(IOTA) for both single electrons and for small countable numbers of
electrons. A reliable and reproducible method to inject or remove a
few electrons was developed. An absolute calibration of the camera
intensity was implemented and its resolving power was evaluated. The
beam lifetime at the lowest intensities was measured. In the central
part of this work, we described how the horizontal, vertical and
longitudinal oscillation amplitudes of a single electron were
measured. From the time evolution of the oscillation amplitudes,
several dynamical quantities were deduced, such as equilibrium
emittances, momentum spread, damping times, and beam energy. The
frequency distribution of residual-gas collisions events vs.\
scattering angle allowed us to estimate residual-gas pressure and
composition and to give an approximate value for the machine aperture.

For the upcoming IOTA experimental runs, we plan to continue this
research, adding to the camera images the synchronized acquisition of
photon arrival times from the photomultipliers. This will allow us to
record phase information for the betatron and synchrotron
oscillations.

These measurements have a general scientific and pedagogic value,
providing direct experimental insights into actual ``single-particle
dynamics'' of an electron in a storage ring. In addition, these
results provide information useful for machine commissioning, for beam
instrumentation and diagnostics, and for verifying ring parameters
obtained with more traditional techniques.

\begin{acknowledgments}
  We would like to thank the entire FAST/IOTA team at Fermilab for
  making these experiments possible, in particular D.~Broemmelsiek,
  K.~Carlson, D.~Crawford, N.~Eddy, D.~Edstrom, R.~Espinoza,
  D.~Franck, V.~Lebedev, S.~Nagaitsev, M.~Obrycki, J.~Ruan, and
  A.~Warner.
  
  This manuscript has been authored by Fermi Research Alliance, LLC
  under Contract No.~DE-AC02-07CH11359 with the U.S.\ Department of
  Energy, Office of Science, Office of High Energy Physics. Work was
  supported in part by the U.S.\ National Science Foundation under
  award PHY-1549132 for the Center for Bright Beams and by the
  University of Chicago.
\end{acknowledgments}

\bibliography{1e_jinst}

\end{document}